\documentclass[prd,nofootinbib,onecolumn,10pt]{revtex4}

\pdfoutput=1

\usepackage{amssymb}
\usepackage{amsmath}
\usepackage{graphicx}
\usepackage{longtable}
\usepackage{verbatim}
\usepackage{amsfonts}
\usepackage{color}
\usepackage[utf8]{inputenc}
\usepackage[normalem]{ulem}
\usepackage[caption=false]{subfig}
\definecolor{BlueViolet}{rgb}{0.21, 0.00, 0.7}
\definecolor{Blue}{rgb}{0.16, 0.00, 0.9}
\definecolor{OGreen}{rgb}{0.33, 0.42, 0.18}
\usepackage[
colorlinks=true,linkcolor=Blue,citecolor=Blue,
urlcolor=BlueViolet,breaklinks=true]{hyperref}

\usepackage{graphicx,amsmath,amssymb,epsfig,verbatim,mathrsfs,array,physics,layout,textcomp,latexsym,xspace,csquotes,placeins,siunitx,soul}
\usepackage{multirow}

\allowdisplaybreaks[1]

\def\s0#1#2{\mbox{\small{$ \frac{#1}{#2} $}}}
\def\0#1#2{\frac{#1}{#2}}
\definecolor{ballblue}{rgb}{0.09, 0.07, 0.9}
\definecolor{darkred}{rgb}{0.8,0.0,0.1}

\newcommand\myb{\mathrel{\stackrel{\makebox[0pt]{\mbox{\normalfont\tiny (--)}}}{b}}}

\begin{document}

\begin{flushright}
	DO-TH 21/07\\
	RBI-ThPhys-2021-13
\end{flushright}

\title{Flavorful leptoquarks at the LHC and beyond: Spin 1}


\author{Gudrun Hiller}
\email{gudrun.hiller@uni-dortmund.de}
\affiliation{Fakult\"at Physik, TU Dortmund, Otto-Hahn-Str.4, D-44221 Dortmund, Germany}
\author{Dennis Loose}
\email{dennis.loose@udo.edu}
\affiliation{Fakult\"at Physik, TU Dortmund, Otto-Hahn-Str.4, D-44221 Dortmund, Germany}
\author{Ivan Ni\v{s}and\v{z}i\'c}
\email{ivan.nisandzic@irb.hr}
\affiliation{Institut Rudjer Bo\v skovi\' c, Division of Theoretical Physics, Bijeni\v cka 54, HR-10000 Zagreb, Croatia}

\begin{abstract}
Evidence for electron-muon universality violation that has been revealed in $b\to s \ell\ell$ transitions  in the observables $R_{K,K^*}$ by the LHCb Collaboration can be explained with  spin-1 leptoquarks
in  $SU(2)_L$  singlet $V_1$ or triplet   $V_3$ representations  in the ${\cal{O}}(1-10)$ TeV range.
We explore the sensitivity of the high luminosity LHC (HL-LHC) and future proton-proton colliders to $V_1$ and $V_3$ in the parameter space 
connected to $R_{K,K^*}$-data.
We consider pair production and single production in association with muons
in different flavor benchmarks.
Reinterpreting a recent ATLAS search for scalar leptoquarks decaying to $b \mu$ and $j \mu$, we extract improved  limits for the leptoquark masses:
For gauge boson-type leptoquarks ($\kappa=1$)  we obtain $M_{V_1}> 1.9$ TeV,  $M_{V_1}> 1.9$ TeV, and $M_{V_1}> 1.7$ TeV 
for leptoquarks decaying predominantly according to
 hierarchical, flipped and democratic  quark flavor structure, respectively.
Future sensitivity projections  based on extrapolations of existing ATLAS and CMS searches are worked out. We find that for $\kappa=1$ the mass reach
for pair (single) production of $V_1$
can be up to 3 TeV (2.1 TeV)  at the HL-LHC and  up to 15 TeV (19.9 TeV) at the FCC-hh with $\sqrt{s}=100$ TeV and $20 \,  \mbox{ab}^{-1}$.
The mass limits and reach for the triplet $V_3$ are similar or higher, depending on flavor.
While there is the exciting possibility that leptoquarks addressing the $R_{K,K^*}$-anomalies are observed at the LHC,
to fully cover the parameter space $pp$-collisions beyond the LHC-energies are needed.
\end{abstract}

\maketitle

\section{Introduction}

The $B$-decay observables $R_K$ and $R_{K^\ast}$~\cite{Hiller:2003js} probe electron-muon universality violating new physics (NP) in 
flavor changing neutral current $b\to s\ell\ell$ transitions.
Recent measurements by the LHCb Collaboration~\cite{Aaij:2019wad, Aaij:2017vbb} returned values that are both about $(15-25)\%$ lower  than the lepton-universality limit $R_K = R_{K^\ast} = 1$, each with statistical significance up to $2.5\,\sigma$.  Very recently, the LHCb Collaboration presented a new update of $R_K$ \cite{Aaij:2021vac} with the central value equal to their previous result \cite{Aaij:2019wad} but with reduced uncertainty, namely: $R_K^{\text{LHCb(2021)}}=0.846^{+0.044}_{-0.041}$, revealing a $3.1\,\sigma$ deviation \cite{Aaij:2021vac} from the SM expectation. Naive combination of $R_K$ and $R_{K^\ast}$ measurements shows a deviation from the standard model above $4\sigma$.  Even more precise measurements are therefore mandatory to clarify these anomalies.
In addition, correlations with other observables are to be expected if lepton universality is actually violated in semileptonic $B$-decays. One important cross check is
 to look for lepton universality breaking 
  in other rare decay channels \cite{Hiller:2014ula}, c.f.\@ recent tests in baryonic decay mode $\Lambda_b \to p K \ell \ell$~\cite{Aaij:2019bzx}, which intriguingly point in the same direction as the ones in $B \to K^{(*)} \ell \ell$  -- a suppression of dimuon modes relative to dielectron ones -- although with lesser significance.
The anomalies also suggest to undertake direct searches 
for the corresponding NP particles at the Large Hadron Collider (LHC) and future high-energy proton-proton ($pp$) colliders~\cite{CEPC-SPPCStudyGroup:2015csa,Zimmermann:2017bbr,Abada:2019ono, Benedikt:2018csr}.

An interesting class of models that can naturally accommodate the measured values of  $R_{K, K^\ast}$ involves a heavy leptoquark that couples to $b\ell$- and $s\ell$ currents, thus contributing to $b \to s \ell\ell$ transitions at tree-level, with masses at scales of up to a few tens of TeV.
Three suitable representations are singled out using both $R_K$ and $R_{K^\ast}$, that is,
the $SU(2)_L$-triplet scalar $S_3 (\bar 3,3,1/3)$, and two spin-1 multiplets, the $SU(2)_L$-singlet $V_1(3, 1, 2/3)$ and the triplet $V_3(3, 3, 2/3)$~\cite{Hiller:2014yaa,Fajfer:2015ycq,Hiller:2014ula,Hiller:2016kry,Hiller:2017bzc,Calibbi:2015kma}.
Specifically, $R_{K}$ and $R_{K^*}$  constrain the leptoquark's coupling to strange quarks and leptons $\ell$, 
$\lambda_{s\ell}$, and
to $b$-quarks and leptons, $\lambda_{b\ell}$, divided by the square of the leptoquark mass $M_V$ as
\begin{equation} \label{eq:V1}
	\frac{ \lambda_{b\mu} \lambda^\ast_{s\mu}- \lambda_{be} \lambda^\ast_{se} }{M_V^2} \simeq - \frac{1\pm 0.24}{ (40\,\text{TeV})^2} \,,
\end{equation}
where we included $R_K$ \cite{Aaij:2021vac} combined with $R_{K^\ast}$\cite{Aaij:2017vbb} in the $q^2$-bin $(1.1, 6)\,\text{GeV}^2$, see Refs.~\cite{Hiller:2014ula, Hiller:2017bzc} for details~\footnote{Current data results in the ratio~\cite{Hiller:2014ula} $X_{K^\ast}\equiv R_{K^\ast}/R_K = 0.84 \pm 0.13$, consistent with $X_{K^\ast}=1$ at $\sim 1.3\,\sigma$, which 
indicates  the possibility of a small admixture from right-handed currents. These could stem from scalar $\tilde S_2(3,2,1/6)$  or vector leptoquarks $V_2(3,2,-5/6)$. In a two-dimensional fit with both left- and right-handed currents the right-hand side of Eq. \eqref{eq:V1} reads $(-1\pm 0.23)/(37\,\text{TeV})^2$.}.
%
One verifies that  {\it i)}  this  indeed requires the leptoquark to couple differently to muons than to electrons, that is, lepton nonuniversality,  {\it ii)} that the data
point to a collider mass scale, and {\it iii)} that  further input is required to extract values of the individual leptoquark couplings, and therefore, the mass.
The latter can be in reach of the LHC if the couplings are sufficiently small.
In general, given sufficient energy, a collider study can not only discover leptoquarks consistent  with (\ref{eq:V1}), but also determine the leptoquark couplings and
mass.

In this paper we focus on the collider signatures of spin-1 (vector) leptoquarks, as a sequel to~\cite{Hiller:2018wbv} on the scalar leptoquark $S_3$.
As in our previous works we assume that the dominant lepton species involved in (\ref{eq:V1}) are muons. This choice is pragmatic, as both lepton species could couple
to NP and sizeable couplings of leptoquarks to electrons are presently not excluded.
However, choosing muons over electrons as the main contributors to  (\ref{eq:V1}) is consistent with the global $b \to s$
fits~\footnote{It is possible that both muon- and electron channels are affected by contributions of opposite sign~\cite{Fornal:2018dqn}.}, which also suggest
NP in $b \to s \mu \mu$ angular distributions.
We stress that dielectron decays $b\to s e e$ also deserve precise experimental treatment in the future \cite{Wehle:2016yoi,Kou:2018nap}.

To quantitatively study the sensitivity to leptoquarks at the LHC and beyond, the leptoquark couplings to bottom- and to strange quarks have to be given individually,
not only their product as in (\ref{eq:V1}).
This is obvious for single production, which feeds on the corresponding parton distribution functions (pdfs) in the proton, see  Fig.~\ref{fig:feynman}, but matters also for pair production \cite{Blumlein:1996qp}, since the flavor patterns
dictate the signature from leptoquark decay.
Flavor symmetries \cite{Froggatt:1978nt} can provide such an input to leptoquark couplings  \cite{Varzielas:2015iva}, implying hierarchical pattern with $\lambda_{s \ell}/\lambda_{b \ell}$ proportional to the strange over the $b$-quark mass. To explore more general settings we employ in addition a flipped benchmark pattern, with inverted hierarchy, and a democratic one.
Note that flavor non-diagonal couplings to leptons and quarks are required  to explain the anomalies (\ref{eq:V1}), as is  taken into account 
in recent leptoquark searches in pair production at ATLAS~\cite{Aad:2020iuy,Aad:2020jmj}; see~\cite{Khachatryan:2015vaa,Khachatryan:2015qda,Sirunyan:2018ryt,Sirunyan:2020zbk} for other  recent  single and pair production searches and \cite{Dorsner:2016wpm} for a review on leptoquark phenomenology.

This paper is organized as follows: In Sec \ref{sec:setup} we introduce the relevant interaction terms for $V_1$ and $V_3$, and introduce three flavor scenarios for the relative size of the couplings to second and third quark generations.
In Sec.~\ref{Collider signatures} we discuss single-, pair- and resonant production mechanisms in the final-state channels involving muons, strange- or bottom quarks 
based on (\ref{eq:V1}) and the flavor benchmarks. We work out present mass limits and determine the sensitivity at the HL-LHC and future $pp$-colliders.
Going beyond  the study  \cite{Hiller:2018wbv} of the $S_3$ leptoquark, here the  projections are  based on extrapolations of existing
LHC searches.
We conclude in Sec.~\ref{Conclusion}. In the appendix we give an approximate analytical argument for comparing $V_1$ and $V_3$ single production cross sections.

\section{Model Setup  \label{sec:setup} }

We briefly review  the vector leptoquarks $V_1$ and $V_3$ and their interactions with standard model particles in Section~\ref{sec:lqs}.
In Section~\ref{Three scenarios} we discuss predictive flavor patterns in the context of the $R_{K^{(*)}}$ anomalies.

\subsection{The vector leptoquarks \texorpdfstring{$V_1$}{V1} and \texorpdfstring{$V_3$}{V3}}
\label{sec:lqs}
We assume that one of the leptoquark representations, either $V_1(3, 1, 2/3)$ or $V_3(3, 3, 2/3)$, provides the resolution of the $R_{K^{(*)}}$ anomalies.

We start by recalling the new interaction terms that need to be added to the SM Lagrangian.
The couplings of $V_1$ to leptons and quarks are
\begin{equation}
	\mathcal L_{\mathrm{int},V_1} = \left(\lambda_{\bar QL} \bar Q \gamma_\mu L + \lambda_{\bar DE} \bar D \gamma_\mu E \right) V_1^\mu + \mathrm{h.c.}\,,\label{Yukawa:V1}
\end{equation}		
while they read for $V_3$ 
\begin{equation}		
	\mathcal L_{\mathrm{int},V_3} = \left(\lambda_{\bar QL} \bar Q \gamma_\mu \vec\sigma L \right) \cdot\vec V_3^\mu + \mathrm{h.c.}\,,\label{Yukawa:V3}
\end{equation}	
where $\vec\sigma$ denotes the Pauli matrices.
After expanding the triplet $V_3$ in terms of its $SU(2)_L$ components
\begin{equation}  \label{eq:SU2-V3}
	\vec\sigma\cdot\vec V_3=
	\begin{pmatrix}
		V_3^{2/3} & \sqrt2 V_3^{5/3} \\
		\sqrt2 V_3^{-1/3} & -V_3^{2/3}
	\end{pmatrix} \,,
\end{equation}
where the superscripts denote the electric charges, the Lagrangian in Eq.~\eqref{Yukawa:V3} can be written as:
\begin{equation}
	\mathcal L_{\mathrm{int},V_3} = -\lambda_{\bar{Q}L}\bar{d}_L\gamma_\mu \ell_L V_3^{2/3\,\mu} + \sqrt{2} \lambda_{\bar{Q}L} \bar{d}_L\gamma_\mu\nu_L V_3^{-1/3\,\mu} + \sqrt{2} \lambda_{\bar{Q}L} \bar{u}_L\gamma_\mu\ell_L V_3^{5/3\,\mu} + \lambda_{\bar{Q}L} \bar{u}_L\gamma_\mu \nu_L V_3^{2/3\,\mu} +\text{h.c.}\label{Eq:Lagrangian-V3}
\end{equation}
The elements of  the coupling matrix $\lambda_{\bar{Q} L}$ for $V_1$ and $V_3$ are  denoted by
\begin{eqnarray}
	\lambda_{\bar{Q} L}=
	\left( \begin{array}{ccc}
			\lambda_{d e} & \lambda_{d \mu} & \lambda_{d \tau}\\
			\lambda_{s e} & \lambda_{s \mu} & \lambda_{s \tau}\\
			\lambda_{b e} & \lambda_{b \mu} & \lambda_{b \tau}\,
	\end{array} \right) \,,\label{Coupling matrix}
\end{eqnarray}
in the mass-basis of the down-type quarks and the charged leptons.

The interactions of $V_1$ and $V_3$ with the standard model gauge bosons follow from the kinetic terms
\begin{equation} \label{eq:V1kin}
	\mathcal L_{\mathrm{kin},V_1} = -\left(D^\mu V_1^\nu \left(D_\mu V_{1\nu}\right)^\dagger - D^\mu V_1^\nu \left(D_\nu V_{1\mu}\right)^\dagger\right) - ig_s\kappa V_1^{\dagger\mu}T^aV_1^\nu G^a_{\mu\nu}-i g_Y \kappa_Y V_1^{\dagger\mu}V_1^\nu B_{\mu\nu}\,,
\end{equation}
and
\begin{equation}
	\mathcal L_{\mathrm{kin},V_3} = -\left(D^\mu \vec V_3^\nu \cdot \left(D_\mu \vec V_{3\nu}\right)^\dagger - D^\mu \vec V_3^\nu \cdot \left(D_\nu \vec V_{3\mu}\right)^\dagger\right) - ig_s\kappa \vec V_3^{\dagger\mu}T^a \cdot \vec V_3^\nu G^a_{\mu\nu} - ig_Y\kappa_Y\vec V_3^{\dagger\mu} \cdot \vec V_3^\nu B_{\mu\nu}-g_2 \kappa_W (\vec V_3^{\dagger\mu}\times \vec V_3^{\nu})\cdot \vec W_{\mu\nu}\,,
\end{equation}
respectively. Here, $T^a$ denote the generators of QCD in the fundamental representation, normalized to ${\rm tr} (T^a T^b)=\delta^{ab}/2$, and $\vec V_3, \vec W_{\mu\nu}$ are the three-component vectors in spin-1 representation of $ SU(2)_L$.
In addition to the terms with covariant derivatives $D^\mu$ renormalizable, gauge invariant interactions with the gluon field strength tensor, parametrized by the coupling $\kappa$,
exist (see e.g.\@ \cite{Rizzo:1996ry, Baker:2019sli}). Other renormalizable couplings of the leptoquark bilinear to weak boson field strengths $W_{\mu\nu}$ and $B_{\mu\nu}$ are irrelevant for the present study.
The value of $\kappa$ depends on the  ultraviolet completion of the model, e.g.\@ $\kappa=1$ in a Yang-Mills case in which the vector leptoquark is the gauge boson of a non-abelian gauge group.
We choose the  benchmark values $\kappa=0$ and $\kappa=1$ throughout the paper, see Sec.~\ref{sec:decay} for a brief discussion of the impact of $\kappa$ on leptoquark production.

We assume that only couplings to quark and lepton doublets are present and hence neglect couplings to singlet fermions in Eq.~\eqref{Yukawa:V1}.
This feature is not generic across possible UV completions and requires some model building, e.g.\@ it does not hold in the minimal Pati-Salam model.
Some of the proposed models in which $V_1$ is a gauge boson also include new vector-like fermions which render the coupling matrix $\lambda_{\bar QL}$ non-unitary.
Several models that can accommodate such a choice for $V_1$ have been proposed in the literature.
For a selection of references studying the $V_1$ leptoquark in the context of the $b\to s \ell\ell$ transitions we refer the reader to Refs.~\cite{Kosnik:2012dj, Hiller:2016kry,Barbieri:2015yvd, DiLuzio:2017vat, Barbieri:2017tuq, Barbieri:2016las, Calibbi:2017qbu, Blanke:2018sro, Fornal:2018dqn, Greljo:2018tuh, Balaji:2019kwe, DiLuzio:2018zxy, Bordone:2017bld, Cornella:2019hct, Fuentes-Martin:2019ign,Angelescu:2018tyl, Bhaskar:2020gkk, Altmannshofer:2020ywf, Dev:2020qet, Mecaj:2020opd, Hati:2020cyn, Bhaskar:2021pml, Crivellin:2021egp,Assad:2017iib}.

\subsection{Three flavor benchmarks}\label{Three scenarios}

The measured values of $R_{K}$~\cite{Aaij:2021vac} and $R_{K^*}$~\cite{Aaij:2017vbb} can be accommodated with 
the combination of couplings and the leptoquark mass given in (\ref{eq:V1}).
It is apparent that an additional constraint on  the leptoquark's  parameter space with regards to collider searches is required.
For instance,  $\lambda_{b\mu} \lambda_{s\mu} \sim 1$ points to a mass scale around \SI{40}{TeV}, outside  of the search range of any presently planned collider, whereas
weaker couplings  $\lambda_{b\mu} \lambda_{s\mu} \sim 10^{-3}-10^{-2}$ imply lower leptoquark mass, in reach of the LHC.
Flavor symmetries, which explain the observed pattern of standard model masses and mixings, do provide naturally requisite suppression mechanisms \cite{Hiller:2016kry}.
These symmetries  determine the ratio between the leptoquark couplings to $b$- and $s$ quarks. We employ these constructions  when
defining flavor benchmark pattern.
In addition,  the $B_s$--$\bar B_s$ mass difference, to which $V_{1,3}$ contribute at 1-loop, combined with $R_{K,K^\ast}$, impose upper bounds of 
around \SI{45}{TeV} and \SI{20}{TeV} on the masses of the $V_1$ and $V_3$ leptoquarks, respectively 
 \cite{Hiller:2017bzc}.  More recent analysis of $B_s$--$\bar B_s$ mixing constraints on NP in $R_{K,K^\ast}$ 
suggest even lower upper mass limits  Ref.~\cite{DiLuzio:2019jyq}. The bound from the loop-induced $B_s$--$\bar B_s$ mixing turns out to be dependent on the specific completion of the vector leptoquark model into a renormalizable theory at high energies, contrary to models with scalar leptoquarks. A comprehensive fit to available leptonic and semileptonic $b\to s \mu\mu$ data sharply supports the $\mathcal{C}_9=-\mathcal{C}_{10}$ solution \cite{Altmannshofer:2021qrr} that corresponds to our models \cite{Hiller:2017bzc} and shows consistency with $R_{K,K^\ast}$ and other muon specific observables. Our scenarios also induce lepton universality violation in the charged current semileptonic B decays within the ratio $R_D^{\mu/e}\equiv \mathcal{B}(B\to D \mu\nu)/\mathcal{B}(B\to D e\nu)$. The measured value \cite{Glattauer:2015teq} $R_D^{\mu/e\,(\text{Belle})}=0.995(22)(39)$  implies the limit $\vert \lambda_{b\mu}\lambda_{s\mu}^\ast\vert/M_{V_{1,3}}^2\lesssim 1/(5.4\,\text{TeV})^2$ which is safely satisfied by our relation \eqref{eq:V1}.
Furthermore, the constraints from perturbative unitarity place an upper bound of about \SI{80} TeV on the scale of effective operator relevant for our present setup~\cite{DiLuzio:2017chi}.

In the following we consider three benchmark scenarios with coupling textures that couple the vector leptoquark predominantly to the second lepton generation.

\begin{description}
	\item[Hierarchical scenario]	
		The first texture we consider is the same as in Ref.~\cite{Hiller:2018wbv} based on flavor models discussed in Ref.~\cite{Hiller:2016kry}, where we assume that the hierarchies found in the standard model masses and mixings  are also present in the leptoquark couplings.
		This is the case in simple flavor models based on the Froggatt-Nielsen-Mechanism~\cite{Froggatt:1978nt} which induces the hierarchies
		\begin{equation}
			\lambda_{d\ell} \ :\ \lambda_{s\ell} \ :\ \lambda_{b\ell} \quad\sim\quad \epsilon^3\dots\epsilon^4 \ :\ \epsilon^2 \ :\ 1
		\end{equation}
		between the different quark generations, where $\epsilon\sim 0.2$ is of the order of the Wolfenstein parameter, i.e. the sine of the Cabibbo angle.
		Specifically, we employ
		\begin{equation}
			\lambda_{\bar QL} \sim \lambda_0 \begin{pmatrix} 0 & 0 & 0 \\ * & \epsilon^2 & * \\ * & 1 & * \end{pmatrix}\,,
			\label{eq:scenario_A}
		\end{equation}
		where we assume contributions to the first quark generation to be suppressed strongly enough to not violate any existing bounds from data on $\mu$-$e$ conversion or rare kaon decays.
		Entries marked with \enquote{$*$} arise only through higher order corrections within the models from Ref.~\cite{Hiller:2016kry}.
		The parametric suppression of the individual quark generations is preserved by Cabibbo-Kobayashi-Maskawa (CKM) rotations.
		As neutrinos are reconstructed inclusively at collider experiments flavor rotations in the lepton sector do not affect such observables.

		Allowing for an additional $\mathcal O(1)$ factor in the ratio between $\lambda_{s\mu}$ and $\lambda_{b\mu}$ couplings, taken  within $1/3$ and $3$, the central value on the right handed side of Eq.~\eqref{eq:V1} implies
			\begin{align} \label{eq:l0range}
			M_{V}/ 14\,\text{TeV} \lesssim \lambda_0 \lesssim M_{V}/ 5\,\text{TeV} \,.
		\end{align}
	These additional flavor model uncertainties  dominate over the   experimental ones in \eqref{eq:V1}.
	\item[Flipped scenario]
		As a second scenario, we consider the inverted form of the previous texture,  that is:
		\begin{equation}
			\lambda_{\bar QL} \sim \lambda_0 \begin{pmatrix} 0 & 0 & 0 \\ * & 1 & * \\ * & \epsilon^2 & * \end{pmatrix}\,.
			\label{eq:scenario_B}
		\end{equation}
		This yields the same effect in $b\to s\ell^+\ell^-$ transitions as the hierarchical pattern while enhancing the single production cross section due to larger pdf  of the strange quark.
We obtain the  same coupling range for $\lambda_0$ as in the hierarchical scenario given in Eq.~\eqref{eq:l0range}.

		Note that this pattern has a weaker foundation in flavor models, and if it is introduced in the interaction basis the CKM rotations can induce contributions to first generation quarks at order $\epsilon$.

	\item[Democratic scenario]
		Lastly, we consider a texture where the couplings to the second and third quark generation are of equal size:
		\begin{equation}
			\lambda_{\bar QL} \sim \lambda_0 \begin{pmatrix} 0 & 0 & 0 \\ * & 1 & * \\ * & 1 & * \end{pmatrix}\,.
			\label{eq:scenario_C}
		\end{equation}
		Taking into account the aforementioned $\mathcal O(1)$ factor and Eq.~\eqref{eq:V1} imply
		\begin{align} \label{eq:l0range_C}
			M_{V}/ 70\,\text{TeV} \lesssim \lambda_0 \lesssim M_{V}/ 23\,\text{TeV} \,.
		\end{align}
\end{description}

Each scenario contains four parameters,  the mass, the parameter $\kappa$ and the dominant couplings $\lambda_{b\mu}$ and $\lambda_{s\mu}$.
The measurements of the single- or pair-production cross section, the corresponding branching fractions and the resonance width, together with the reconstruction of the mass peak, would suffice to determine all four parameters. Note that $b$-tagging would be necessary for such an analysis.

We stress that flavor models which isolate a single species of leptons are straightforward to obtain using techniques from neutrino model building~ \cite{Varzielas:2015iva},
however, more general settings can arise and are viable, too.
Allowing for significant entries \enquote{$*$} in (\ref{eq:scenario_A}), (\ref{eq:scenario_B}), (\ref{eq:scenario_C}) would open up further leptoquark decay modes and search channels, which  would
reduce  branching ratios in the signal channels studied here.
Negligible entries \enquote{$*$}  correspond therefore to the most favorable situation for an observation in the muon channel.

Leptoquarks $V_{1,3}$ and $S_3$ contribute also at tree level to charged current $b \to c \ell \nu$-decays \cite{Fajfer:2012jt, Fajfer:2015ycq, Hiller:2016kry, Barbieri:2015yvd, DiLuzio:2017vat, Barbieri:2017tuq, Barbieri:2016las, Calibbi:2017qbu, Blanke:2018sro, Greljo:2018tuh, Balaji:2019kwe, DiLuzio:2018zxy, Bordone:2017bld, Cornella:2019hct, Fuentes-Martin:2019ign,Angelescu:2018tyl}.
In particular $V_1$, which evades dominant constraints from $b \to s \nu \bar \nu$ on third generation lepton couplings, has received interest
as a possible resolution of the anomalies in the  $b \to c \tau \nu$ observables  $R_{D^{(*)}}$. Note that \eqref{eq:V1} points to a NP contribution to
a  loop-induced process in the standard model, hence a corresponding NP effect in tree level charged currents would naturally be 
${\cal{O}}(1/(4 \pi)^2 \cdot \lambda_{ q \tau}/\lambda_{q \mu})$-suppressed. An effect of the same order of magnitude in the charged current as in the neutral current, which
is about  ten percent, would therefore require substantial hierarchy $\lambda_{q \tau} \sim 10^2 \lambda_{q \mu}$, which is unsupported by flavor models and points to strong couplings to $\tau$'s.
As a study in concrete, full flavor models \cite{Hiller:2016kry} is beyond the scope of this work, we also do not consider links with the  $b \to c \tau \nu$ observables.

\section{Collider phenomenology}\label{Collider signatures}

In this section we study vector leptoquark production in $pp$-collisions and decays of leptoquarks. Basics are given in Sec.~\ref{sec:decay}.
We work out  bounds on the masses of vector leptoquarks using  available search results for pair-production of leptoquarks from ATLAS~\cite{Aad:2020iuy}
(Sec.~\ref{Current bounds}),
and cross sections for  future $pp$ colliders in Sec.~\ref{sec:future}.
We consider three  setups corresponding to center-of-mass energies $\sqrt{s}$: $14\,\mathrm{TeV}$ (LHC run 3), $27\,\mathrm{TeV}$ (HE-LHC), and $100\,\mathrm{TeV}$ (FCC-hh)~\cite{Zimmermann:2017bbr} with target integrated luminosities of $\mathcal L=\SI{3}{\per\atto\barn}$, \SI{15}{\per\atto\barn} and \SI{20}{\per\atto\barn}, respectively. In Sec.~\ref{sec:resonant} we also briefly discuss resonant production.
We analyze the mass reach of future $pp$ colliders by extrapolating current limits on cross sections to higher center-of-mass 
energies and luminosities   in Sec.~\ref{Sec:Synopsis}.

\subsection{Leptoquark production and decay \label{sec:decay}}

\begin{figure}
	\centering
	\begin{tabular}{cccc}
		\multicolumn{2}{c}{\quad\quad\quad\subfloat[]{\includegraphics[height=0.12\textwidth]{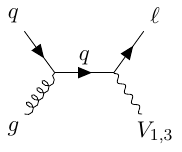}}} &
		\multicolumn{2}{c}{\subfloat[]{\includegraphics[height=0.12\textwidth]{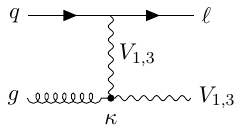}}} \\
		\subfloat[]{\includegraphics[height=0.12\textwidth]{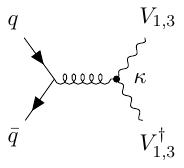}} &
		\multicolumn{2}{c}{\quad\quad\subfloat[]{\includegraphics[height=0.10\textwidth]{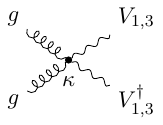}}} &
		\subfloat[]{\includegraphics[height=0.10\textwidth]{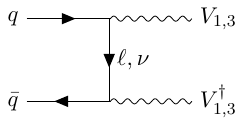}}\\
		\subfloat[]{\includegraphics[height=0.12\textwidth]{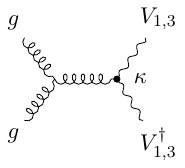}} &
		\multicolumn{2}{c}{\quad\quad\subfloat[]{\includegraphics[height=0.12\textwidth]{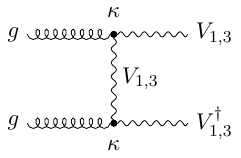}}} &
		\subfloat[]{\includegraphics[height=0.12\textwidth]{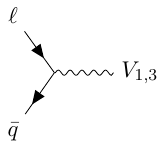}}
	\end{tabular}
	\caption{Leading order Feynman diagrams for  single production, (a) and (b), pair production, (c) - (g), and resonant production, (h), of the $V_1$ and $V_3$ vector leptoquarks.
		The dots indicate additional contributions stemming from the coupling with the gluon field strength tensor proportional to $\kappa$.
		Notice the crossed versions of diagram (g), which are not shown explicitly. 
			}
	\label{fig:feynman}
\end{figure}

We consider three dominant mechanisms of leptoquark production at $pp$ colliders: pair production, single production in association with a lepton and resonant-production induced by quark-lepton fusion, shown  in Fig.~\ref{fig:feynman}.
The structure of the final state signatures of all three classes of processes is determined by the flavor structure of the leptoquark couplings.

The flavor scenarios  (\ref{eq:scenario_A}), (\ref{eq:scenario_B}), (\ref{eq:scenario_C}) can be  distinguished experimentally 
by different patterns of the final states in two-body decays of the leptoquarks.
In the hierarchical scenario  (\ref{eq:scenario_A}), the dominant leptoquark decay modes are
\begin{equation} \label{eq:hiV1}
	V_1^{+2/3} \to b\mu^+\,,\ t\bar\nu\,,
\end{equation}
for the singlet and
\begin{equation} \label{eq:hiV3}
	\begin{aligned}
		V_3^{-1/3} &\to b\bar\nu \,,\\
		V_3^{+2/3} &\to b\mu^+\,,\ t\bar\nu \,,\\
		V_3^{+5/3} &\to t\mu^+\,,
	\end{aligned}
\end{equation}
for the triplet.
The $b\mu^+$ and $t\bar\nu$ final states of the $V_{1,3}^{+2/3}$ leptoquarks are related by  $SU(2)_L$ symmetry such that their branching fractions are approximately equal.

In the flipped scenario  (\ref{eq:scenario_B}) the leading signatures involve charm and strange quarks 
\begin{equation}
	V_1^{+2/3} \to s\mu^+\,,\ c\bar\nu\,,
\end{equation}
for the singlet and
\begin{equation}
	\begin{aligned}
		V_3^{-1/3} &\to s\bar\nu\,, \\
		V_3^{+2/3} &\to s\mu^+\,,\ c\bar\nu\,, \\
		V_3^{+5/3} &\to c\mu^+\,,
	\end{aligned}
\end{equation}
for the triplet.
In the democratic scenario (\ref{eq:scenario_C})  all of the above modes arise  and final states with both light and heavy quarks are relevant.
We recall that we allow in (\ref{eq:scenario_C})  a mild hierarchy between the two couplings which  can have a strong impact on the relative size of the different final state branching ratios as $\mathcal B \sim |\lambda_{\bar Q \ell}|^2$.  Approximate branching ratios  for the  benchmark patterns  (\ref{eq:scenario_A}), (\ref{eq:scenario_B}), (\ref{eq:scenario_C})  are given in Tables \ref{tab:flavor_BRs_V1} and \ref{tab:flavor_BRs_II}.

\begin{table}[h]
	\centering
	\begin{tabular}{m{3cm}m{1cm}m{1cm}m{1cm}m{1cm}}
		\hline
		& $b\mu^+$ & $t\bar \nu$ & $s\mu^+$ & $c\bar \nu$ \\
		\hline
		hierarchical& $1/2$ & $1/2$ & $0$ & $0$ \\
		flipped & $0$ & $0$ & $1/2$ & $1/2$ \\
		democratic & $1/4$ & $1/4$ & $1/4$ & $1/4$ \\
		\hline
	\end{tabular}
	\caption{Branching fractions of the $V_1$ leptoquark and the triplet component $V_3^{2/3}$ in the benchmark scenarios from Sec.~\ref{Three scenarios}. }
	\label{tab:flavor_BRs_V1}
\end{table}

\begin{table}[h]
	\centering
	\begin{tabular}{m{3cm}m{1.5cm}m{1.5cm}}
		\hline
		& $b\bar\nu\enskip (t\mu^+)$ & $s\bar\nu\enskip (c\mu^+)$ \\
		\hline
		hierarchical & $1$ & $0$ \\
		flipped & $0$ & $1$ \\
		democratic & $1/2$ & $1/2$ \\
		\hline
	\end{tabular}
	\caption{Branching fractions of the $V_{3}^{-1/3}$ ($V_{3}^{+5/3}$) leptoquarks in the benchmark scenarios from Sec.~\ref{Three scenarios}.}
	\label{tab:flavor_BRs_II}
\end{table}

\begin{figure}
	\begin{center}
		\includegraphics[height=0.3\textwidth]{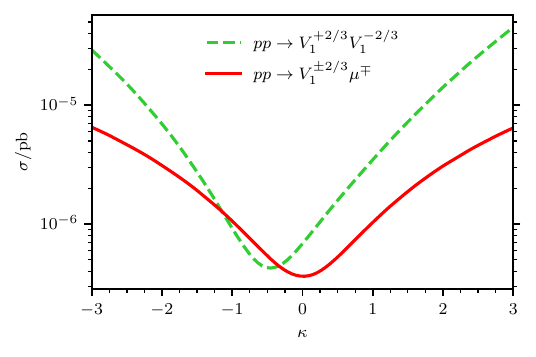}
	\end{center}
	\caption{$\kappa$-dependence (\ref{eq:V1kin}) of the single- (red, solid) and pair production cross section (green, dashed) for $V_1$.
		We fix $\sqrt s=\SI{14}{TeV}$ and $M_{V_1}=\SI{3}{TeV}$.
		For the single production cross section we employ the hierarchical scenario.
		Analogous results are obtained for other choices of the parameters and flavor benchmarks.
	}
	\label{fig:kappa_dep}
\end{figure}

In Fig.~\ref{fig:kappa_dep} we show  the pair- and single production cross sections as functions of $\kappa$  for the example of 
the HL-LHC $\sqrt s=\SI{14}{TeV}$ and $M_{V_1}=\SI{3}{TeV}$.
The cross sections exhibit  minima for  $\kappa$ in the vicinity of $0$ or $-1$, see also~\cite{Rizzo:1996ry}. The shapes vary mildly with the variation of the leptoquark mass
 in  a  range suitable for a  \SI{14} TeV collider.

\subsection{Current mass bounds}\label{Current bounds}

Leptoquark-based explanations of the deviations found in B-physics motivated the recent search for pair-produced scalar leptoquarks with $139\, \text{fb}^{-1}$ of data from \SI{13}{TeV} $p\text{-}p$ collisions~\cite{Aad:2020jmj}.
Mass limits for scalar leptoquarks decaying dominantly to top and electron (muon), obtained from this search, are \SI{1470}{GeV} (\SI{1480}{GeV}).
Such final states appear in decays of scalar (vector) $SU(2)_L$ triplet leptoquark $S_3$ ($V_3$).
Previously derived collider bounds on vector leptoquarks from pair production searches are $M_{V_1} > \SI{1.3}{TeV}$ for the dominant decays to $\tau b$, and $M_{V_1} > \SI{1.7}{TeV}$ in the $\mu b$ channel~\cite{Diaz:2017lit}, for $\kappa=1$. 

\begin{figure}
	\begin{center}
		{\includegraphics[width=0.46\textwidth]{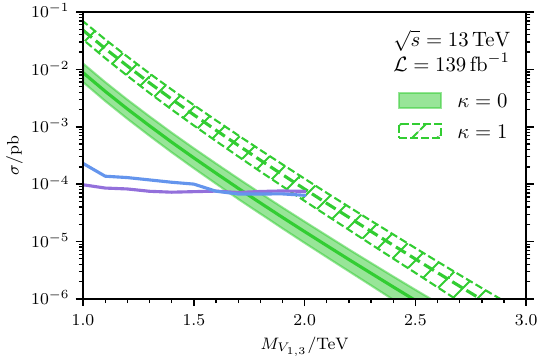}}
		{\includegraphics[width=0.46\textwidth]{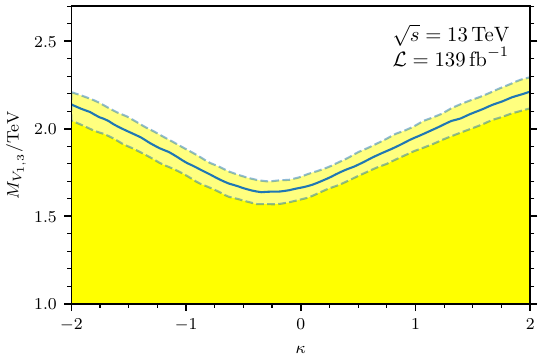}}
		{\includegraphics[width=0.46\textwidth]{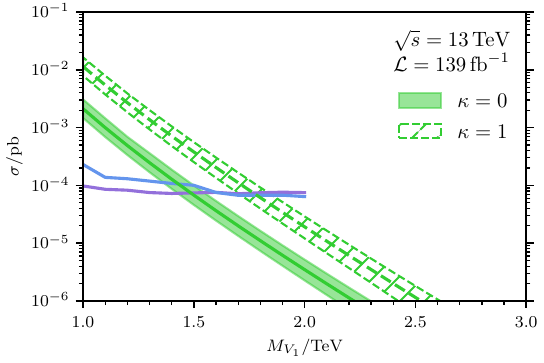}}
		{\includegraphics[width=0.46\textwidth]{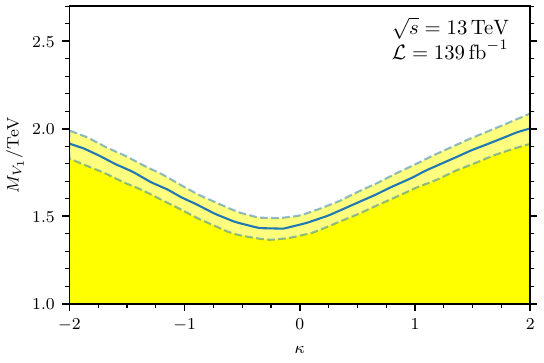}}
		{\includegraphics[width=0.46\textwidth]{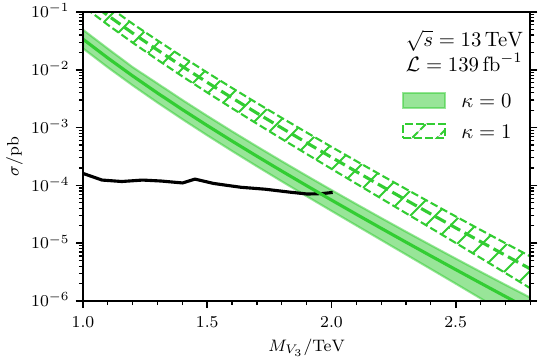}}
		{\includegraphics[width=0.46\textwidth]{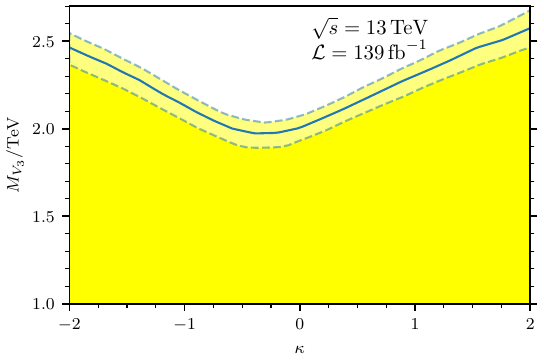}}
	\end{center}
		\caption{Sensitivity and mass bounds from reinterpretation of a current ATLAS search~\cite{Aad:2020iuy}. Top row:  $V_1$ in  the hierarchical and flipped flavor scenarios, which equals $V_3$ in  the hierarchical and democratic flavor scenarios,
		middle row: $V_1$ in the democratic scenario, and bottom row: $V_3$ in the flipped scenario.
		Left: Sensitivity to  $V_1,V_3$-pair production, assuming dominant decays to $\mu b$, $\mu q$ and $\mu c$  in purple, blue and black, respectively; $q$ denotes quarks lighter than the  charm quark,
		The green bands indicate the theory prediction including the pdf- and scale uncertainties, for $\kappa=0$ (solid) and $\kappa=1$ (dashed).
		Right: Mass bounds for the leptoquarks $V_1,V_3$ 
		as the function of parameter $\kappa$  (\ref{eq:V1kin}).
		The boundary of the excluded region is represented by the band whose width results from the pdf- and scale uncertainties.}
	\label{Fig:atlas}
\end{figure}

We evaluate the current mass limits for the $V_1$ and $V_3$ leptoquarks using the limits on the cross sections found in the ATLAS Collaboration search~\cite{Aad:2020iuy} for  pair production of scalar leptoquarks.
This search was performed using the data collected in the \SI{13}{TeV} LHC run with a luminosity of $\mathcal{L}=\SI{139}{\per\femto\barn}$.
For the hierarchical scenario we use the limits obtained in the $(b\mu\,, b\mu)$-channel while for the flipped scenario we use the ($q\mu\,, q\mu$)-channel. The role of $q$ in the latter channel is played in our  $V_1$ model by the strange quark,  and by charm for $V_3$. The bound in the democratic scenario is obtained from the $(b\mu\,, b\mu)$-channel. 
The limits on the cross section in the $(b \mu,\, b \mu )$-, $(q\mu, q\mu)$-  and $(c \mu, c \mu)$-channels (bottom plot) are shown in Fig.~\ref{Fig:atlas} (plots to the left) -- the comparison to the theoretical cross sections for leptoquark pair production and subsequent decay in corresponding final-state channels determines the mass limits.  

We now spell out the obtained limits for the $V_1$-leptoquark model. For both the hierarchical and the flipped scenarios the limits are \SI{1.7(1)}{TeV} and \SI{2.0(1)}{TeV} for $\kappa = 0$ and $\kappa = 1$, respectively. The bounds are the same  for both of these scenarios because the current experimental limits for the cross sections to $(b\mu, b\mu)$- and $(s\mu,s\mu)$ final states nearly coincide in the region of large leptoquark masses. The limits are somewhat weaker for the case of democratic scenario and read \SI{1.5(1)}{TeV} and \SI{1.8(1)}{TeV} for $\kappa = 0$ and $\kappa = 1$, due to smaller individual branching fractions into $b\mu$- and $s\mu$-pairs, see Tab~\ref{tab:flavor_BRs_V1}. For the democratic scenario, in which the leptoquark
couplings to $b-$ and $s$ quarks are approximately equal and both of the above mentioned limits apply, we used $(b\mu, b\mu)$-channel, since the corresponding experimental limit is currently somewhat more strict in the most of the explored mass range than the one for $(q\mu, q\mu)$.

In the case of the model with the $V_3$ leptoquark in the hierarchical scenario the limit is the same as for the case of $V_1$: \SI{1.7(1)}{TeV} and \SI{2.0(1)}{TeV}, where the role of the $V_1^{2/3}$ state is now played by the triplet component $V_3^{2/3}$. In the flipped scenario, the respective limits are stronger: \SI{2.0(1)}{TeV} and \SI{2.3(1)}{TeV}. This is due to the $V_3^{-5/3}V_3^{+5/3}$-pair contributing to the final states $(c\mu, c\mu)$  with large branching fractions
${\cal{B}}(V_3^{5/3}\to c\mu^+)\sim 1$, see Tab.~\ref{tab:flavor_BRs_II}. The corresponding final state has been included in the search by the ATLAS Collaboration~\cite{Aad:2020iuy}. This channel becomes the leading one in the determination of the bound for the democratic scenario as well, resulting in \SI{1.7(1)}{TeV} and \SI{2.0(1)}{TeV} for $\kappa = 0$ and $\kappa = 1$, respectively. 

The mass limits depend on the value of $\kappa$, as shown in Fig.~\ref{Fig:atlas} (plots to the right).
Note that the cross sections have a minimum within $\kappa \in (-2,2)$, which corresponds to the weakest bound on the mass. 
These are the same for $V_1$ and $V_3$ in the hierarchical scenario, $M_{V_{1,3}}>1.6\,\text{TeV}$ for $\kappa=-0.3$,
and equal to one for $V_1$ with flipped and $V_3$ with democratic flavor structure.
 Note, however, that the absolute minimum for the bound on the $V_1$ mass is obtained in the democratic scenario and reads $M_{V_1}>1.4\,\text{TeV}$ for $\kappa=-0.3$. The corresponding weakest bound in the case of $V_3$ is found in the case of hierarchical scenario, given above.

Constraints on the parameter space of leptoquark models can also be obtained from Drell-Yan processes, to which the leptoquarks contribute via  $t$-channel exchange~\cite{Faroughy:2016osc, Greljo:2017vvb}. For our present setup, the corresponding di-muon channel is relevant. The authors of Refs.~\cite{Greljo:2017vvb, Angelescu:2018tyl} performed the recast of the ATLAS collaboration measurement~\cite{Aaboud:2017buh} in the dimuon channel at \SI{13}{TeV} with $36\,\text{fb}^{-1}$ of data, while the authors of Ref.~\cite{Bhaskar:2021pml} used the result by the CMS collaboration \cite{Sirunyan:2018ipj} corresponding to the same center-of-mass energy and luminosity. The results of these studies imply that the parameter space relevant for our model does not receive constraints at present. This situation could change as more data is collected in the future~\cite{Greljo:2017vvb}. 

We find that the cross sections for  pair production of the $V_1$ leptoquark typically turn out several times larger than those for any specific weak-isospin component of the scalar leptoquark $S_3$ studied in Ref.~\cite{Hiller:2018wbv}. This is not surprising, given that the vector leptoquark involves three helicity states. Thus, for given values of  the branching fractions in the specific lepton-quark channels, taken  to be equal for the scalar and the vector leptoquark, the corresponding search limits for the scalar turn out weaker, in accord with what was previously noted in Ref.~\cite{Angelescu:2018tyl}. However, we note that the pattern of the branching fractions into the final state lepton-quark pairs is guided by the details of the flavor structure of the Yukawa couplings to the fermions, and the $SU(2)_L$ structure of a leptoquark representation. For example,
for $M_{\text{LQ}}=3\,\text{TeV}$ and $\sqrt{s}=14\,\text{TeV}$, we have $\sigma(pp\to S_3^{4/3}S_3^{-4/3})=10^{-7}\text{pb}$ and $\sigma(pp\to V_1^{+2/3}V_1^{-2/3})=6.9\cdot 10^{-7}\text{pb}$. However, within {\it e.g.} the  hierarchical scenario in Eq.~\eqref{eq:scenario_A} we have $\mathcal{B}(S_3^{4/3}\to b\mu)\simeq 1$, while $\mathcal{B}(V_1^{2/3}\to b\mu)\simeq 1/2$, which lowers the cross section of the vector pair-production in the $(b\mu, b\mu)$ channel by factor $1/4$.

\subsection{Single and pair production cross sections \label{sec:future}}

\begin{figure}
	\centering
	\includegraphics[height=0.93\textwidth]{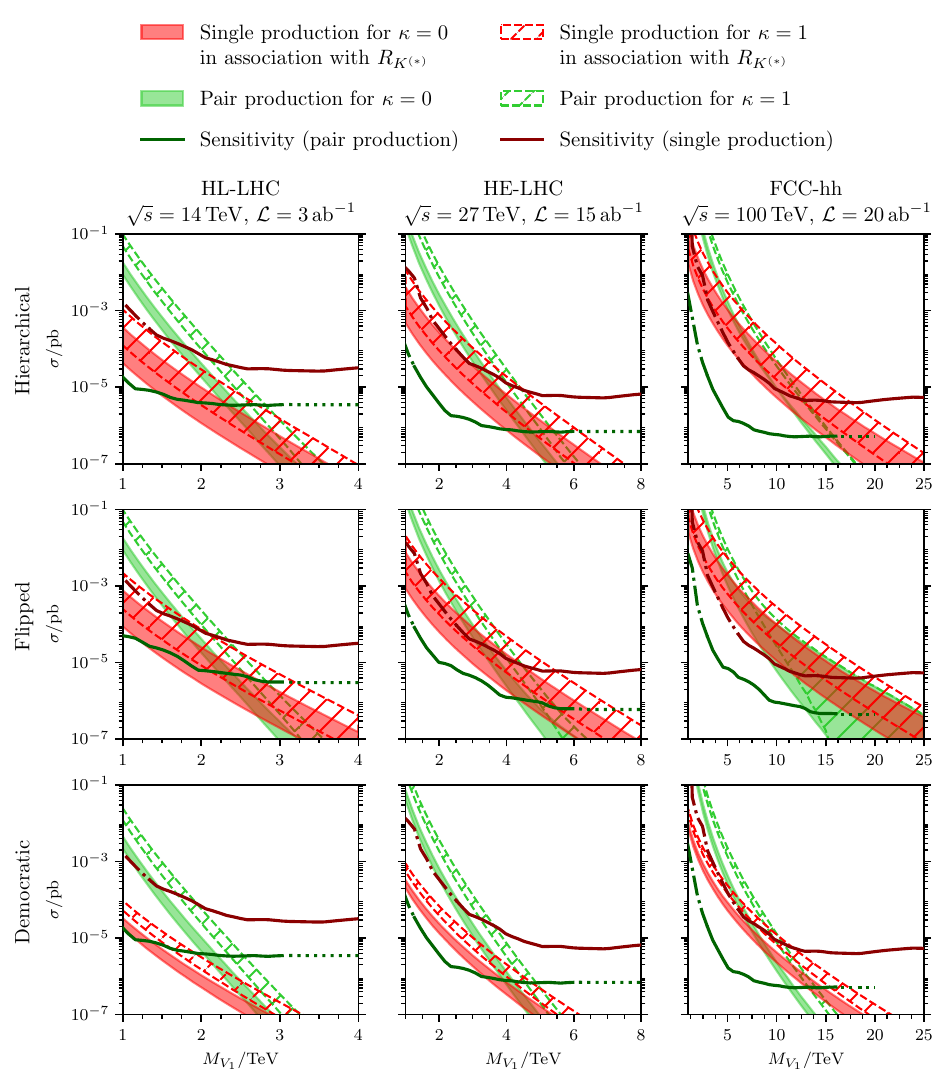}
	\caption{$V_1$-leptoquark production in $pp$-collisions in the  flavor scenarios introduced in Sec.~\ref{Three scenarios} (rows) for different future collider experiments (columns).
		Red bands: Single production cross section for $\sigma(p p\to V_1^{\pm 2/3}( \to \mu^{\pm}\myb)\mu^\mp)+\sigma(pp\to V_1^{\pm 2/3}( \to \mu^{\pm}j)\mu^\mp)$, derived from the $R_{K,K^*}$-band in Eq. \eqref{eq:l0range} for hierarchical and flipped scenarios, and Eq.~\eqref{eq:l0range_C} for the democratic scenario. Light green: pair production with final states $(b\mu, b\mu)$ for the hierarchical and democratic scenarios and $(q\mu, q\mu)$ for the flipped scenario. The error bands for pair production are evaluated by combining the pdf-, and scale uncertainties.
		Results for $\kappa=1$ are shown in a dashed/hatched form together with the solid curves for $\kappa=0$.
		The solid dark red and the dark green curves depict the projected experimental sensitivity for single and pair production, respectively. As the starting curves for these extrapolations, we used the results of the measurements by the CMS~\cite{Khachatryan:2015qda} and ATLAS~\cite{Aad:2020iuy} collaborations, for the single- and pair production, respectively, see Sec.~\ref{Sec:Synopsis} for details. The dot-dashed segments of the extrapolated curves for the low masses required additional smooth variation of the luminosities between the initial and the target values, following the prescription in Ref.~\cite{Allanach:2017bta}. The dotted segments for the large masses represent the smooth continuation above the final extrapolated points towards the higher masses with the constant values of the cross-section limit. Neither of these segments play a role in determining the mass reaches shown in Tab.~\ref{tab:mass_reach}.}
	\label{fig:production-mu}
\end{figure}

\begin{figure}[h]
	\begin{center}
		\includegraphics[height=0.93\textwidth]{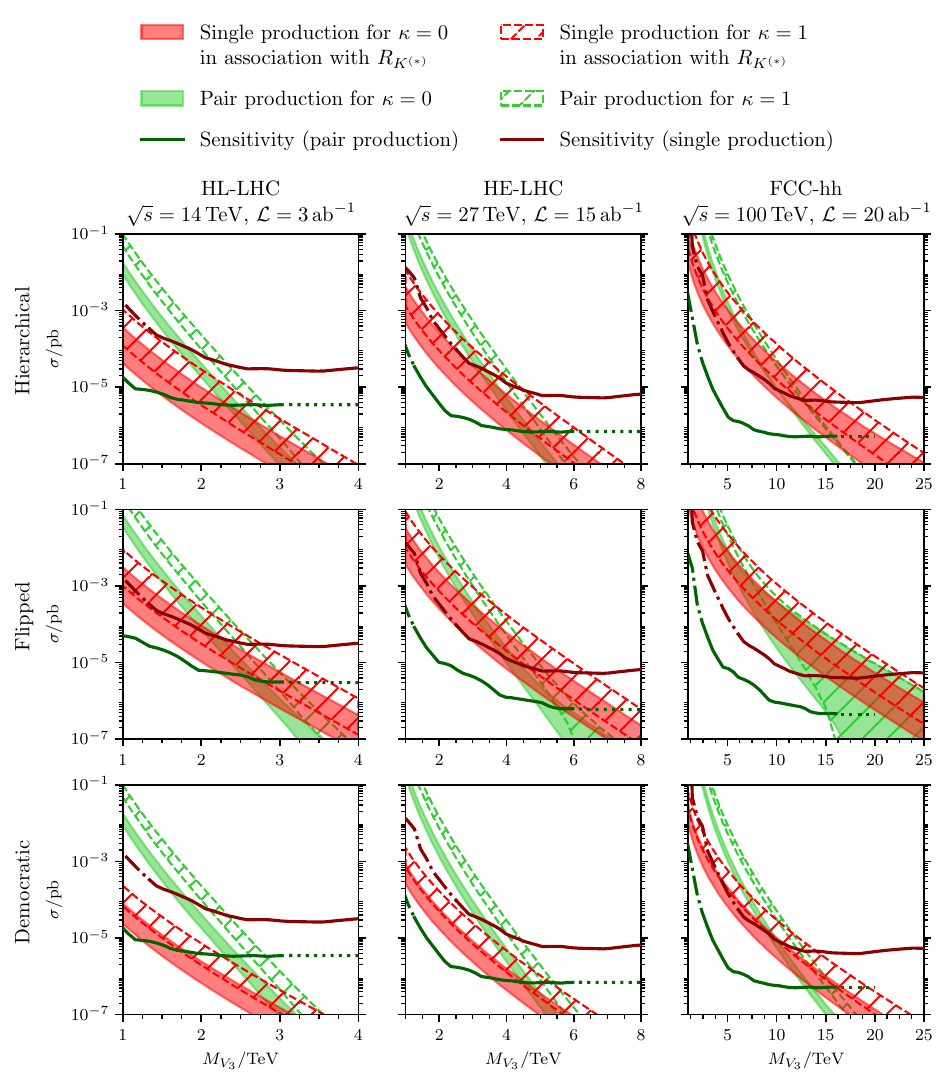}
	\end{center}
	\caption{$V_3$-leptoquark production in $pp$-collisions in the flavor scenarios introduced in Sec.~\ref{Three scenarios} (rows) for different future collider experiments (columns).
		Red bands: Single production cross section for $\sigma(p p\to \mu^+\mu^- j)$ induced by the triplet $V_3$, derived from the $R_{K,K^*}$-band in Eq. \eqref{eq:l0range} for hierarchical and flipped scenarios, and Eq.~\eqref{eq:l0range_C} for the democratic scenario. Light green: pair production with final states $(b\mu, b\mu)$ for the hierarchical and democratic scenarios and $(c\mu, c\mu)$ for the flipped scenario. The error bands for pair production are evaluated by combining the pdf-, and scale uncertainties, see Fig.~\ref{fig:production-mu} and Sec.~\ref{Sec:Synopsis} for the details. }
	\label{fig:mass_scan_plot_V3}
\end{figure}

We evaluate the leading order cross sections for the single production of $V_1$ in association with a muon, represented by the resonant diagrams (a) and (b) in Fig.~\ref{fig:feynman}, as functions of the leptoquark mass.
The results are displayed by  the red bands using $\kappa=0$ in Fig.~\ref{fig:production-mu} for the case of $V_1$, and   in Fig.~\ref{fig:mass_scan_plot_V3} for $V_3$.  The results for $\kappa=1$ are displayed by hatched bands.
The leading-order cross sections for pair production and subsequent resonant decays are represented by the solid (hatched) light green bands
for $\kappa=0$ $(\kappa=1)$.

For each flavor scenario we assume that the parameters of the leptoquark model satisfy Eq.~\eqref{eq:V1}.
The band widths originate from  Eqs.~\eqref{eq:l0range}, \eqref{eq:l0range_C}.
We note that there are also  non-resonant diagrams contributing to $(q\ell\,, q\ell)$ final states that were not taken into account in our numerical analysis, assuming that the contributions of the resonant diagrams shown Fig.~\ref{fig:feynman} (c)-(g) are well separated by  appropriate kinematic cuts.

For the evaluation of the cross sections and the corresponding uncertainty bands we used \texttt{Madgraph}~\cite{Alwall:2014hca} with the \texttt{UFO} \cite{Degrande:2011ua} output of the leptoquark models that we implemented using \texttt{Feynrules}~\cite{Alloul:2013bka}.
The pdf-, and scale uncertainties are evaluated using \texttt{LHAPDF}~\cite{Buckley:2014ana}, symmetrized and combined in quadrature.
Our \texttt{Feynrules} implementations of the $V_1$ and $V_3$ models are attached to this paper as ancillary files.
We checked the consistency of our \texttt{Feynrules} implementations with the corresponding implementations from Ref.~\cite{Dorsner:2018ynv}. We used the software package \texttt{Feyncalc}~\cite{Shtabovenko:2020gxv,Shtabovenko:2016sxi} for several cross-checks.

Pair production is predominantly induced by the QCD-initiated processes and is essentially independent of the flavor structure.
The latter determines the branching fractions into various final state channels, see Tab.~\ref{tab:flavor_BRs_V1}.
As an exception to this, the large contribution of diagram (e) shown in Fig.~\ref{fig:feynman} becomes  noticeable for the large-mass region within the flipped scenario, see the last plot of the second row in Fig.~\ref{fig:production-mu}.
In this case, the $R_{K, K^\ast}$-condition \eqref{eq:V1} for flipped flavor hierarchy, forces the large values for the coupling $\lambda_{s\mu}$, reaching the borders of the perturbativity. This as well as the larger uncertainty band in this plot can be understood from \eqref{eq:l0range} for large values of $M_V$.

The magnitude of the single production cross section induced by $q g \to V_1^{2/3} \ell$ at parton level is directly proportional to the square of the magnitude of the corresponding flavor coupling $\lambda_{\bar{Q}\ell}$. Assuming the narrow width approximation, we multiply the corresponding production cross sections by the corresponding branching fractions given in Tab.~\ref{tab:flavor_BRs_V1}. Since there are no available single production searches involving $b$ quarks in the final state, we added the contributions involving jets and $b$-quarks which amounts to the branching fraction $1/2$ for each of the three flavor scenarios.  

\FloatBarrier
\subsection{Resonant production \label{sec:resonant}}

Determinations of the photon distribution function inside the proton introduced in Refs.~\cite{Manohar:2016nzj,Manohar:2017eqh} were recently followed by the determination of the lepton pdfs in Ref.~\cite{Buonocore:2020nai}.
These results opened  up the possibility to  consider  resonant leptoquark production from lepton-quark fusion in $pp$ collisions~\cite{Buonocore:2020erb}, 
see diagram (h) in Fig.~\ref{fig:feynman}.
Next-to-leading-order QCD and QED corrections to the resonant production of scalar leptoquarks have recently become available~\cite{Greljo:2020tgv}.

\begin{figure}
	\begin{center}
		\includegraphics[height=0.38\textwidth]{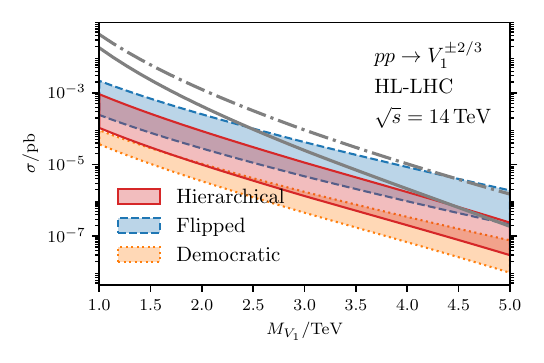}
	\end{center}
	\caption{
		Resonant leptoquark production cross section from lepton-quark fusion for the flavor scenarios  
		(\ref{eq:scenario_A}), (\ref{eq:scenario_B}), (\ref{eq:scenario_C})  at the HL-LHC.
		The solid (dash-dotted) grey line indicates the resonant cross section with only the $b\mu$ ($s\mu$) coupling set to one.
}
	\label{fig:resonant_production}
\end{figure}

To illustrate the expected range  within the flavor scenarios  (\ref{eq:scenario_A}), (\ref{eq:scenario_B}), (\ref{eq:scenario_C}), we give in
Fig.~\ref{fig:resonant_production} the  resonant  cross section for  $V_1$   at a $\sqrt{s}=\SI{14}{TeV}$ $pp$ collider.
The results are obtained by convolution of the leading order partonic cross section for $\mu\,(b+s) \to V_1$ with the \texttt{LUXlep-NNPDF31\_nlo\_as\_0118\_luxqed}~\cite{Buonocore:2020nai} pdf set that includes the leptonic pdfs.
Note that the charge conjugated process is also included in the results.
We parse the pdf set in \texttt{Mathematica} using the package \texttt{ManeParse}~\cite{Clark:2016jgm}.
The resulting cross sections are larger than those of pair- and single production due to lesser phase-space suppression.
It would be desirable to look for collider signatures of this process.
Resonant vector leptoquark collider signatures motivated by the $R_{D, D^\ast}$ anomalies, and the corresponding backgrounds, were recently discussed in Ref.~\cite{Haisch:2020xjd}.

In Fig.~\ref{fig:sqrts_vs_mass} we compare the cross sections of the resonant- and single  production for $V_1$-masses up to $\sim 10\,\text{TeV}$. The regions of the $(M_{V_1},\sqrt{s}$)-plane  to the right of  the thick blue lines result in resonant cross section being  larger than the one for single production. The current level of the lepton pdf-uncertainties does not allow for  extrapolations to higher energy scales.

We note in  passing the absence of the triplet leptoquark component $V_3^{-1/3}$ in the resonant production -- its coupling to the fermion sector exclusively involves neutrinos, see Eq.\eqref{Eq:Lagrangian-V3}.

\subsection{Sensitivity projections for future colliders}\label{Sec:Synopsis}

In order to estimate the mass reach of the future colliders for the flavor benchmark scenarios, we extrapolate existing bounds from single- and pair production using the limit extrapolation method following Refs.~\cite{Thamm:2015zwa, Allanach:2017bta}.
The method assumes that the exclusion limits are determined by the numbers of background events and involves the appropriate re-scaling of the background processes with the corresponding parton luminosity functions, see \cite{Thamm:2015zwa, Allanach:2017bta} for more details.
We expect the method to be less suitable for the case of leptoquarks than for {\it e.g.,}\@ the case of $s$-channel resonances, for which it was initially used~\cite{Thamm:2015zwa}, however, it should provide the correct estimate of the order of magnitude for the collider limits on the corresponding cross sections. 

As the starting point for our approximation for the future sensitivity projections for single production, we employ  the limits obtained by the CMS Collaboration in the $\sqrt{s} = \SI{8}{TeV}$ run 
with  $\mathcal{L}=19.6~\text{fb}^{-1}$~\cite{Khachatryan:2015qda}.
The latter paper  presents the limits on the resonant cross sections of the single production of the leptoquarks in  association with muons in the $\mu\mu j$ final states.
Our extrapolations assume that the final $b$ quark is not tagged and is counted as a light jet, however, we stress once again that  $b$-tagging is required for distinguishing between the flavor scenarios, and could lead to improved limits in the case of the hierarchical and democratic flavor scenarios.

For the extrapolations of the limits on cross sections for pair production we use the search performed by the ATLAS Collaboration in Ref.~\cite{Aad:2020iuy} at \SI{13}{TeV} with $\mathcal{L}=\SI{139}{\per\femto\barn}$, see also Refs.~\cite{Sirunyan:2018ryt,Khachatryan:2015vaa} for earlier searches.
We use the leading order set of pdfs provided by MSTW Collaboration~\cite{Martin:2009iq} for the evaluation of the extrapolations.
As a cross-check, we use the pdf-set \texttt{NNPDF23\_lo\_as\_0130\_qed} \cite{Ball:2013hta}, parsed using the package \texttt{ManeParse}~\cite{Clark:2016jgm}.
For the hierarchical and democratic scenarios we extrapolate the limits in the $(b\mu\,, b\mu)$-channel, while the limits for the $(q\mu\,, q\mu)$-channel were used for the flipped scenario, where the role of $q$ is played by the strange quark.

The extrapolations of the limits for the single- and pair production cross sections are compared to the corresponding theoretical resonant cross sections for  $V_1$ in Fig.~\ref{fig:production-mu}. The comparison for the case of $V_3$ is given in Fig.~\ref{fig:mass_scan_plot_V3}.
We find good agreement with the similar extrapolations for the case of single-production in Ref.~\cite{Allanach:2017bta} .\footnote{Up to date analysis of the future sensitivity for the pair production of scalar leptoquarks were recently presented in Ref.~\cite{Allanach:2019zfr}.}

We provide a list of possible  mass reaches at  future colliders for the leptoquark $V_1$, in each of the three flavor scenarios, in Tab.~\ref{tab:mass_reach}, for both pair- and single production channels.
The reach for $V_3$ is given separately  in parentheses, if different from the reach in $V_1$.
\begin{table}
	\centering
	\begin{tabular}{c|c|c|c|c|c|c|c|c|c|c}
		\hline
		\multirow{2}{*}{Collider} & \multirow{2}{*}{$\sqrt s/\si{TeV}$} & \multirow{2}{*}{$\mathcal{L}\,/\si{\per\atto\barn}$} & \multicolumn{4}{c|}{Mass reach for $\kappa=0$} & \multicolumn{4}{c}{Mass reach for $\kappa=1$} \\
		\cline{4-11}
		& & & hierarchical & flipped & democratic & pair & hierarchical & flipped & democratic & pair \\
		\cline{4-11}
		\hline
		HL-LHC & 14 & 3 & $\text{---}$ & $(2.3)$ & $\text{---}$ & $2\ (3)$ & $\text{---}$ & $2.1\ (2.8)$ & $\text{---}$ & $3\ (3)$ \\
		HE-LHC & 27 & 15 & $2.7$ & $4.4\ (5.6)$ & $\text{---}$ & $5\ (5)$ & $4.5$ & $5.5\ (6.4)$ & $\text{---}$ & $5\ (6)$ \\
		FCC-hh & 100 & 20 & $15.1$ & $17.7\ (20.5)$ & $(10.7)$ & $13\ (15)$ & $17.5$ & $19.9\ (22.7)$ & $11.7\ (14.0)$ & $15\ (18)$ \\
		\hline
	\end{tabular}
	\caption{
		Mass reach in TeV for vector leptoquark single production in the hierarchical, flipped and democratic scenarios from Sec.~\ref{Three scenarios} and pair production, at different future colliders for  $\kappa=0$ and $\kappa=1$.
		For  single production we provide the mass reaches corresponding to the upper limit of the cross section band resulting from  Eqs.~(\ref{eq:l0range}), (\ref{eq:l0range_C}).
					In the flipped and democratic scenarios as well as for  pair production we show the increased mass reaches for $V_3$ in parentheses, for the hierarchical scenario the $V_1$,$V_3$ reaches are the same,  see Appendix~\ref{Appendix A} for details.
	}
	\label{tab:mass_reach}
\end{table}

\begin{figure}
	\begin{center}
		\includegraphics[height=0.53\textwidth]{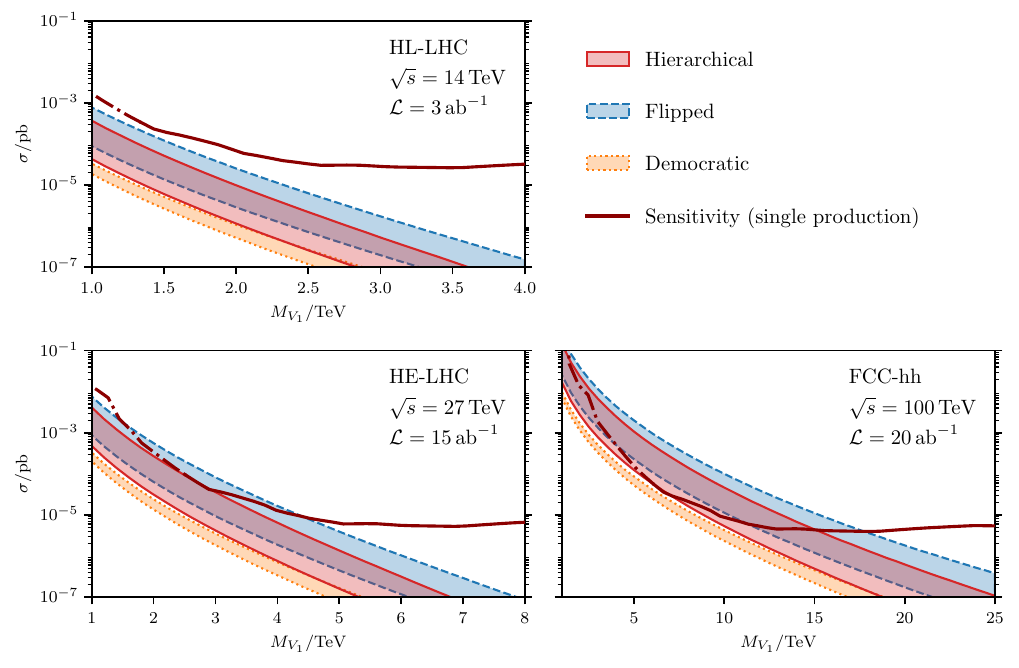}
	\end{center}
	\caption{Comparison of the single leptoquark production cross sections $\sigma(pp\to V_1^{\pm 2/3}\mu^\mp)$ for the benchmarks (\ref{eq:scenario_A}), (\ref{eq:scenario_B}), (\ref{eq:scenario_C})  at different future colliders, for $\kappa=0$.
	}
	\label{fig:mass_scan_plot_combined}
\end{figure}

As can be seen from Figs.~\ref{fig:production-mu} and \ref{fig:mass_scan_plot_V3}  the theoretical predictions for  single production in association with muons at \SI{14}{TeV} and \SI{27}{TeV} colliders turn out to be rather small, below the  projected sensitivity.
In Fig.~\ref{fig:mass_scan_plot_combined} we compare the expectations for the single production cross sections for the  flavor scenarios and different future collider experiments; the values span up to two orders of magnitude.

Observation of a single production signal with a cross section that is much larger than those shown in Fig.~\ref{fig:mass_scan_plot_combined} would point to a leptoquark that is unrelated to $R_{K, K^\ast}$, since the simultaneous leptoquark couplings to all three flavors $(d,s,b)$ are  restricted by the low energy flavor-changing-neutral current (FCNC) observables, such as kaon decays.

\begin{figure}
	\begin{center}
		\includegraphics[height=0.5\textwidth]{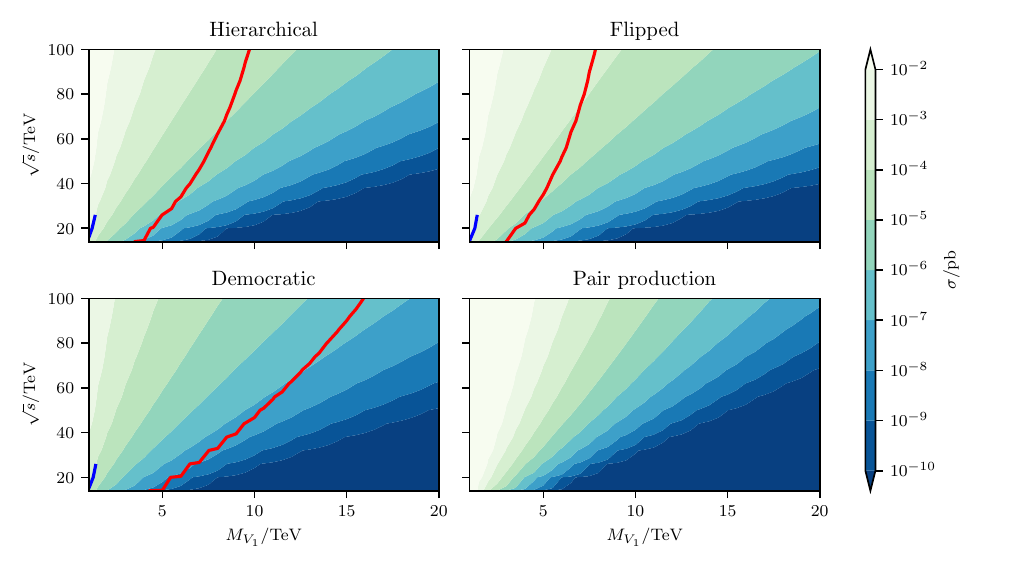}
	\end{center}
	\caption{Single leptoquark production cross section for $V_1$  depending on the center-of-mass energy $\sqrt s$, and the leptoquark mass.
		For each scenario we use the central value of the allowed ranges from Eqs.~\eqref{eq:l0range}, \eqref{eq:l0range_C}.
	In the regions to the right of the red lines the single leptoquark production cross section is larger than the pair production cross section.
	In the regions to the right of the blue lines (up to $M_{V_1}\sim 10\,\text{TeV}$) the resonant leptoquark production cross section is larger than the single production one, see text.
	In the plot to the lower right we show in addition the pair production cross section $\sigma (pp \to V_1^{+2/3}  V^{-2/3}_1)$. All plots are for $\kappa=0$.}
	\label{fig:sqrts_vs_mass}
\end{figure}

In Fig.~\ref{fig:sqrts_vs_mass} we compare the cross sections for single- and pair production in the $(M_V, \sqrt{s})$ plane, where $\sqrt{s}$ denotes the center-of-mass energy of the $p\text{-}p$ collisions.
Pair production is instrumental for the discovery or the exclusion of vector leptoquarks in the region of a few TeV.
For large masses and scattering energies, the cross sections of the single production turn out larger than those of the pair production -- the corresponding regions are located on the right of the solid red lines in the first three plots of the Fig.~\ref{fig:mass_scan_plot_combined}.
Notice that the cross sections of the single production vary significantly with the different flavor scenarios.
In case of a signal discovery, $b$-tagging would be important in order to confirm the connection to the $R_{K, K^\ast}$-anomalies.

\FloatBarrier
\section{Conclusions}\label{Conclusion}

Leptoquarks are flavorful -- a feature that  allows for rich phenomenology and model-dependence alike.
While the recent evidence reported by the LHCb Collaboration~\cite{Aaij:2021vac} for the breakdown of lepton universality in rare semileptonic $b$-decays has yet to be confirmed by experiments and in other observables,
taking the data at face value provides informative directions in the leptoquarks' parameter space:
Among the spin 1 leptoquark representations only $V_1$ and $V_3$ induce sufficiently large contributions explaining $R_{K,K^*}$ at tree-level \cite{Hiller:2017bzc}, with
coupling over mass ratio fixed (\ref{eq:V1}). 
Here we study the vector leptoquark reach  at the LHC and beyond targeting this parameter space.
Specifically, we analyze signatures from leptoquarks with couplings to second and third generation quark doublets, and to muons. The reason  why  this
simplified framework is sensible is two-fold:
There is presently no necessity to consider couplings to electrons, and flavor symmetries explaining neutrino masses and mixing result in
leptoquark couplings to a single lepton species \cite{Varzielas:2015iva}. We stress that  dedicated searches for leptoquarks decaying to leptons other
than muons \cite{Aad:2020jmj} are well-motivated and complementary, however, beyond the scope of this work.

We work out single- and pair production cross sections in three quark flavor benchmarks: a hierarchical one  (\ref{eq:scenario_A}), with dominant coupling to third generation quarks, a flipped one  (\ref{eq:scenario_B}), with dominant coupling to second  generation quarks, and a democratic one (\ref{eq:scenario_C}).
Reinterpreting a recent ATLAS search for pair-produced scalar leptoquarks~\cite{Aad:2020iuy}
we obtain the mass limit $M_{V_1}> 1.4$ TeV for $\kappa=-0.3$, and higher otherwise. Analogous limits can be derived for $V_3$, neglecting mass splitting within the multiplet,  as $M_{V_3}>1.6\,\text{TeV}$ for $\kappa=-0.3$. Limits for gauge-type leptoquarks ($\kappa=1$), or without the $\kappa$-term are stronger, see Sec.~\ref{Current bounds}.

The future reach  at the HL-LHC, with $\sqrt{s}=14$ TeV and $3\,\mbox{ab}^{-1}$, the HE-LHC with 27 TeV and $15 \mbox{ab}^{-1}$ and the FCC-hh with 100 TeV
and $20\,\mbox{ab}^{-1}$ is shown in Figs.~\ref{fig:production-mu} and \ref{fig:mass_scan_plot_V3}, and summarized in Table \ref{tab:mass_reach}.  
For $\kappa=1$, the maximal reach for $V_1$ is  3 TeV  (HL-LHC), $5.5$ TeV (HE-LHC) and $19.9$ TeV (FCC-hh). The reach for the triplet $V_3$ is similar 
in single production and the hierarchical pattern, and generically larger otherwise.
All cross sections become larger for larger value of the  parameter $| \kappa|$, as illustrated in Fig.~\ref{fig:kappa_dep}, and improve the mass reach.
Results are based on extrapolations of CMS~\cite{Khachatryan:2015qda} and ATLAS \cite{Aad:2020iuy} searches. Pair production has larger cross sections
due to the strong interaction  until  phase space suppression kicks in and single production takes over, as demonstrated  quantitatively in  Fig.~\ref{fig:sqrts_vs_mass}.
Single production is, however, valuable on its own  as it is sensitive to  the flavor patterns in the new physics sector. The flipped pattern with subject to the larger pdf gives largest
cross sections, followed by the hierarchical one, see also Fig.~\ref{fig:mass_scan_plot_combined}.  

Recent works suggest to study resonant production from lepton-quark fusion at the LHC via lepton pdfs. Similar to single production the cross section
is sensitive to flavor, as shown in Fig.~\ref{fig:resonant_production} for the HL-LHC. A full study of efficiencies also at future machines is beyond the scope of this work.

Note also that the patterns  (\ref{eq:scenario_A}), (\ref{eq:scenario_B}), (\ref{eq:scenario_C}) are simplified and in general
 lepton flavor violating signatures can arise in leptoquark decays, {\it e.g.,}  \cite{Hiller:2017bzc}.
Allowing for significant entries $(\ast)$ in the patterns would open up further search channels, and reduces leptoquark branching ratios in the signal channels studied here.
Note that this rescaling effect is linear in single production, and quadratic in pair production, but  leaves the qualitative features, such as quark flavor hierarchies of our analysis
intact. A study in concrete, full flavor models \cite{Hiller:2016kry} is beyond the scope of this work.

We conclude that leptoquark searches at the LHC are very well motivated by flavor physics, although covering  the full mass range supported presently by 
rare processes requires higher energies. The actual confirmation of the $R_{K,K^*}$-anomalies would strengthen the case for a corresponding machine.
Observing leptoquarks directly would disentangle mass from couplings, as in (\ref{eq:V1}), and could distinguish flavor patterns. Besides being striking signals from beyond the standard model, this would allow to make progress
 towards the flavor puzzle. 
 
\section*{Acknowledgements}
I.N. acknowledges support provided by the Alexander von Humboldt Foundation within the framework of the Research Group Linkage Programme funded by the German Federal Ministry of Education and Research.

\appendix
\section{Comparing  \texorpdfstring{$V_3$}{V3} to $V_1$ production}\label{Appendix A}

The cross sections of $V_3$ are larger than the ones of $V_1$ in the flipped and democratic scenario, due to the contributions from the 
additional components in the $SU(2)_L$ triplet. Here we give analytical arguments for  the approximate relations between the cross sections of single production.
In the hierarchical flavor scenario the dominant contribution is from $b$-quarks which involves
only the $V_3^{2/3}$ from the triplet \eqref{eq:hiV3} which makes the cross section equal to the singlet \eqref{eq:hiV1} one. 

In order to estimate the cross section  of singly produced $V_3$ with signature $pp\to j\mu\mu$  in the flipped and democratic scenarios, we use $\sigma(pp\to V_3^{2/3}\mu^-) = \sigma(pp\to V_1^{2/3}\mu^-)$ and include the contribution from $cg\to V_3^{5/3}(\to j\mu^+)\mu^-$.
In the flipped scenario holds
\begin{equation}
	\sigma_{V_3}^\text{Flipped}(pp\to j\mu\mu) = 2\big[\sigma(sg\to V_3^{2/3}\mu^-)\mathcal B(V_3^{2/3}\to s\mu^+)
	+ \sigma(cg\to V_3^{5/3}\mu^-)\mathcal B(V_3^{5/3}\to c\mu^+)\big]\,,
\end{equation}
where the factor of $2$ stem from adding the CP-conjugate of the process.
Assuming that the pdfs for the strange and charm quarks are roughly the same, one obtains $\sigma(sg\to V_3^{2/3}\mu^-)\simeq(\sqrt 2)^2\,\sigma(cg\to V_3^{5/3}\mu^-)$, 
where the $(\sqrt 2)^2$ is an isospin factor \eqref{eq:SU2-V3}.
Using the different branching ratios given in Tables \ref{tab:flavor_BRs_V1} and \ref{tab:flavor_BRs_II}, we find $\sigma_{V_3}^\text{Flipped}(pp\to j\mu\mu) \simeq 5\,\sigma_{V_1}^\text{Flipped}(pp\to j\mu\mu)$. An explicit numerical evaluation reveals the ratio $\sigma_{V_3}^\text{Flipped}(pp\to j\mu\mu)/\sigma_{V_1}^\text{Flipped}(pp\to j\mu\mu)$ to be decreasing for larger leptoquark mass, and within 4 and $2.5$ for values of $(\sqrt{s}, M_V)$  in the range of interest for the present paper.
For the democratic scenario we find
\begin{equation}
	\begin{split}
		\sigma_{V_3}^\text{Democratic}(pp\to j\mu\mu) &= 2\big[\sigma(bg\to V_3^{2/3}\mu^-)\mathcal B(V_3^{2/3}\to (b,s)\mu^+) + \sigma(sg\to V_3^{2/3}\mu^-)\mathcal B(V_3^{2/3}\to (b,s)\mu^+) \\
		&+ \sigma(cg\to V_3^{5/3}\mu^-)\mathcal B(V_3^{5/3}\to c\mu^+)\big] \\
	\end{split}
\end{equation}
Assuming $\sigma(sg\to V_3^{2/3}\mu^-)=4\,\sigma(bg\to V_3^{2/3}\mu^-)$ results in $\sigma_{V_3}^\text{Democratic}(pp\to j\mu\mu) \simeq 2.5\,\sigma_{V_1}^\text{Democratic}(pp\to j\mu\mu)$, whereas an  explicit evaluation results the ratio to be $2.5$, and dropping for larger masses to  $1.5$, for relevant ranges of  $(\sqrt{s}, M_V)$. Numerical results from the explicit evaluations of $\sigma(cg\to V_3^{5/3}\mu^-)$ are  included in Fig.~\ref{fig:mass_scan_plot_V3}.

\FloatBarrier
\providecommand{\href}[2]{#2}
\begingroup\raggedright


\begin{thebibliography}{99}


\bibitem{Hiller:2003js}
G.~Hiller and F.~Kr\"uger,
\href{http://dx.doi.org/doi:10.1103/PhysRevD.69.074020}{
Phys. Rev. D \textbf{69}, 074020 (2004)}
[arXiv:hep-ph/0310219 [hep-ph]].

\bibitem{Aaij:2019wad}
R.~Aaij \textit{et al.} [LHCb],
\href{http://dx.doi.org/doi:10.1103/PhysRevLett.122.191801}{
Phys. Rev. Lett. \textbf{122}, no.19, 191801 (2019)}
[arXiv:1903.09252 [hep-ex]].

\bibitem{Aaij:2017vbb}
R.~Aaij \textit{et al.} [LHCb],
\href{http://dx.doi.org/doi:10.1007/JHEP08(2017)055}{
JHEP \textbf{08}, 055 (2017)}
[arXiv:1705.05802 [hep-ex]].

\bibitem{Aaij:2021vac}
R.~Aaij \textit{et al.} [LHCb],
[arXiv:2103.11769 [hep-ex]].

\bibitem{Hiller:2014ula}
G.~Hiller and M.~Schmaltz,
\href{http://dx.doi.org/doi:10.1007/JHEP02(2015)055}{
JHEP \textbf{02}, 055 (2015)}
[arXiv:1411.4773 [hep-ph]].

\bibitem{Aaij:2019bzx}
R.~Aaij \textit{et al.} [LHCb],
\href{http://dx.doi.org/doi:10.1007/JHEP05(2020)040}{
JHEP \textbf{05}, 040 (2020)}
[arXiv:1912.08139 [hep-ex]].

\bibitem{CEPC-SPPCStudyGroup:2015csa}
M.~Ahmad, D.~Alves, H.~An, Q.~An, A.~Arhrib, N.~Arkani-Hamed, I.~Ahmed, Y.~Bai, R.~B.~Ferroli and Y.~Ban, \textit{et al.}
IHEP-CEPC-DR-2015-01.

\bibitem{Zimmermann:2017bbr}
F.~Zimmermann,
ICFA Beam Dyn. Newslett. \textbf{72}, 138-141 (2017)

\bibitem{Abada:2019ono}
A.~Abada \textit{et al.} [FCC],
\href{http://dx.doi.org/doi:10.1140/epjst/e2019-900088-6}{
Eur. Phys. J. ST \textbf{228}, no.5, 1109-1382 (2019)}

\bibitem{Benedikt:2018csr}
A.~Abada \textit{et al.} [FCC],
\href{http://dx.doi.org/doi:10.1140/epjst/e2019-900087-0}{
Eur. Phys. J. ST \textbf{228}, no.4, 755-1107 (2019)}

\bibitem{Hiller:2014yaa}
G.~Hiller and M.~Schmaltz,
\href{http://dx.doi.org/doi:10.1103/PhysRevD.90.054014}{
Phys. Rev. D \textbf{90}, 054014 (2014)}
[arXiv:1408.1627 [hep-ph]].

\bibitem{Fajfer:2015ycq}
S.~Fajfer and N.~Ko\v{s}nik,
\href{http://dx.doi.org/doi:10.1016/j.physletb.2016.02.018}{
Phys. Lett. B \textbf{755}, 270-274 (2016)}
[arXiv:1511.06024 [hep-ph]].

\bibitem{Hiller:2016kry}
G.~Hiller, D.~Loose and K.~Sch\"onwald,
\href{http://dx.doi.org/doi:10.1007/JHEP12(2016)027}{
JHEP \textbf{12}, 027 (2016)}
[arXiv:1609.08895 [hep-ph]].

\bibitem{Hiller:2017bzc}
G.~Hiller and I.~Nisandzic,
\href{http://dx.doi.org/doi:10.1103/PhysRevD.96.035003}{
Phys. Rev. D \textbf{96}, no.3, 035003 (2017)}
[arXiv:1704.05444 [hep-ph]].

\bibitem{Calibbi:2015kma}
L.~Calibbi, A.~Crivellin and T.~Ota,
\href{http://dx.doi.org/doi:10.1103/PhysRevLett.115.181801}
{Phys. Rev. Lett. \textbf{115} (2015), 181801}
[arXiv:1506.02661 [hep-ph]].

\bibitem{Hiller:2018wbv}
G.~Hiller, D.~Loose and I.~Ni\v{s}and\v{z}i\'c,
\href{http://dx.doi.org/doi:10.1103/PhysRevD.97.075004}{
Phys. Rev. D \textbf{97}, no.7, 075004 (2018)}
[arXiv:1801.09399 [hep-ph]].



\bibitem{Wehle:2016yoi}
S.~Wehle \textit{et al.} [Belle],
\href{http://dx.doi.org/doi:10.1103/PhysRevLett.118.111801}{
Phys. Rev. Lett. \textbf{118}, no.11, 111801 (2017)}
[arXiv:1612.05014 [hep-ex]].

\bibitem{Kou:2018nap}
E.~Kou \textit{et al.} [Belle-II],
PTEP \textbf{2019}, no.12, 123C01 (2019)
\href{http://dx.doi.org/doi:10.1093/ptep/ptz106}{
[erratum: PTEP \textbf{2020}, no.2, 029201 (2020)]}
[arXiv:1808.10567 [hep-ex]].

\bibitem{Blumlein:1996qp}
J.~Bl\"umlein, E.~Boos and A.~Kryukov,
\href{http://dx.doi.org/doi:10.1007/s002880050538}{Z. Phys. C \textbf{76} (1997), 137-153}
[arXiv:hep-ph/9610408 [hep-ph]].

\bibitem{Froggatt:1978nt}
C.~D.~Froggatt and H.~B.~Nielsen,
\href{http://dx.doi.org/doi:10.1016/0550-3213(79)90316-X}{
Nucl. Phys. B \textbf{147}, 277-298 (1979)}

\bibitem{Varzielas:2015iva}
I.~de Medeiros Varzielas and G.~Hiller,
\href{http://dx.doi.org/doi:10.1007/JHEP06(2015)072}{
JHEP \textbf{06}, 072 (2015)}
[arXiv:1503.01084 [hep-ph]].

\bibitem{Aad:2020iuy}
G.~Aad \textit{et al.} [ATLAS],
\href{http://dx.doi.org/doi:10.1007/JHEP10(2020)112}{
JHEP \textbf{10}, 112 (2020)}
[arXiv:2006.05872 [hep-ex]].

\bibitem{Aad:2020jmj}
G.~Aad \textit{et al.} [ATLAS],
[arXiv:2010.02098 [hep-ex]].

\bibitem{Khachatryan:2015vaa}
V.~Khachatryan \textit{et al.} [CMS],
\href{http://dx.doi.org/doi:10.1103/PhysRevD.93.032004}{
Phys. Rev. D \textbf{93}, no.3, 032004 (2016)}
[arXiv:1509.03744 [hep-ex]].

\bibitem{Khachatryan:2015qda}
V.~Khachatryan \textit{et al.} [CMS],
Phys. Rev. D \textbf{93}, no.3, 032005 (2016)
\href{http://dx.doi.org/doi:10.1103/PhysRevD.93.032005}{
[erratum: Phys. Rev. D \textbf{95}, no.3, 039906 (2017)]}
[arXiv:1509.03750 [hep-ex]].

\bibitem{Sirunyan:2018ryt}
A.~M.~Sirunyan \textit{et al.} [CMS],
\href{http://dx.doi.org/doi:10.1103/PhysRevD.99.032014}{
Phys. Rev. D \textbf{99}, no.3, 032014 (2019)}
[arXiv:1808.05082 [hep-ex]].

\bibitem{Sirunyan:2020zbk}
A.~M.~Sirunyan \textit{et al.} [CMS],
[arXiv:2012.04178 [hep-ex]].

\bibitem{Dorsner:2016wpm}
I.~Dor\v{s}ner, S.~Fajfer, A.~Greljo, J.~F.~Kamenik and N.~Ko\v{s}nik,
\href{http://dx.doi.org/doi:10.1016/j.physrep.2016.06.001}{
Phys. Rept. \textbf{641}, 1-68 (2016)}
[arXiv:1603.04993 [hep-ph]].

\bibitem{Rizzo:1996ry}
T.~G.~Rizzo,
eConf \textbf{C960625}, NEW151 (1996)
[arXiv:hep-ph/9609267 [hep-ph]].

\bibitem{Baker:2019sli}
M.~J.~Baker, J.~Fuentes-Mart\'\i{}n, G.~Isidori and M.~K\"onig,
\href{http://dx.doi.org/doi:doi:10.1140/epjc/s10052-019-6853-x}{Eur. Phys. J. C \textbf{79} (2019) no.4, 334}
[arXiv:1901.10480 [hep-ph]].

\bibitem{Kosnik:2012dj}
N.~Kosnik,
\href{http://dx.doi.org/doi:10.1103/PhysRevD.86.055004}{
Phys. Rev. D \textbf{86}, 055004 (2012)}
[arXiv:1206.2970 [hep-ph]].

\bibitem{Barbieri:2015yvd}
R.~Barbieri, G.~Isidori, A.~Pattori and F.~Senia,
\href{http://dx.doi.org/doi:10.1140/epjc/s10052-016-3905-3}{
Eur. Phys. J. C \textbf{76}, no.2, 67 (2016)}
[arXiv:1512.01560 [hep-ph]].

\bibitem{DiLuzio:2017vat}
L.~Di Luzio, A.~Greljo and M.~Nardecchia,
\href{http://dx.doi.org/doi:10.1103/PhysRevD.96.115011}{
Phys. Rev. D \textbf{96}, no.11, 115011 (2017)}
[arXiv:1708.08450 [hep-ph]].

\bibitem{Barbieri:2017tuq}
R.~Barbieri and A.~Tesi,
\href{http://dx.doi.org/doi:10.1140/epjc/s10052-018-5680-9}{
Eur. Phys. J. C \textbf{78}, no.3, 193 (2018)}
[arXiv:1712.06844 [hep-ph]].

\bibitem{Barbieri:2016las}
R.~Barbieri, C.~W.~Murphy and F.~Senia,
\href{http://dx.doi.org/doi:10.1140/epjc/s10052-016-4578-7}{
Eur. Phys. J. C \textbf{77}, no.1, 8 (2017)}
[arXiv:1611.04930 [hep-ph]].

\bibitem{Calibbi:2017qbu}
L.~Calibbi, A.~Crivellin and T.~Li,
\href{http://dx.doi.org/doi:10.1103/PhysRevD.98.115002}{
Phys. Rev. D \textbf{98}, no.11, 115002 (2018)}
[arXiv:1709.00692 [hep-ph]].

\bibitem{Blanke:2018sro}
M.~Blanke and A.~Crivellin,
\href{http://dx.doi.org/doi:10.1103/PhysRevLett.121.011801}{
Phys. Rev. Lett. \textbf{121}, no.1, 011801 (2018)}
[arXiv:1801.07256 [hep-ph]].

\bibitem{Greljo:2018tuh}
A.~Greljo and B.~A.~Stefanek,
\href{http://dx.doi.org/doi:10.1016/j.physletb.2018.05.033}{
Phys. Lett. B \textbf{782}, 131-138 (2018)}
[arXiv:1802.04274 [hep-ph]].

\bibitem{Balaji:2019kwe}
S.~Balaji and M.~A.~Schmidt,
\href{http://dx.doi.org/doi:10.1103/PhysRevD.101.015026}{
Phys. Rev. D \textbf{101}, no.1, 015026 (2020)}
[arXiv:1911.08873 [hep-ph]].

\bibitem{DiLuzio:2018zxy}
L.~Di Luzio, J.~Fuentes-Martin, A.~Greljo, M.~Nardecchia and S.~Renner,
\href{http://dx.doi.org/doi:10.1007/JHEP11(2018)081}{
JHEP \textbf{11}, 081 (2018)}
[arXiv:1808.00942 [hep-ph]].

\bibitem{Bordone:2017bld}
M.~Bordone, C.~Cornella, J.~Fuentes-Martin and G.~Isidori,
\href{http://dx.doi.org/doi:10.1016/j.physletb.2018.02.011}{
Phys. Lett. B \textbf{779}, 317-323 (2018)}
[arXiv:1712.01368 [hep-ph]].

\bibitem{Cornella:2019hct}
C.~Cornella, J.~Fuentes-Martin and G.~Isidori,
\href{http://dx.doi.org/doi:10.1007/JHEP07(2019)168}{
JHEP \textbf{07}, 168 (2019)}
[arXiv:1903.11517 [hep-ph]].

\bibitem{Fuentes-Martin:2019ign}
J.~Fuentes-Mart\'\i{}n, G.~Isidori, M.~K\"onig and N.~Selimovi\'c,
\href{http://dx.doi.org/doi:10.1103/PhysRevD.101.035024}{
Phys. Rev. D \textbf{101}, no.3, 035024 (2020)}
[arXiv:1910.13474 [hep-ph]].

\bibitem{Angelescu:2018tyl}
A.~Angelescu, D.~Be\v{c}irevi\'c, D.~A.~Faroughy and O.~Sumensari,
\href{http://dx.doi.org/doi:10.1007/JHEP10(2018)183}{
JHEP \textbf{10}, 183 (2018)}
[arXiv:1808.08179 [hep-ph]].

\bibitem{Bhaskar:2020gkk}
A.~Bhaskar, T.~Mandal and S.~Mitra,
\href{http://dx.doi.org/doi:10.1103/PhysRevD.101.115015}{
Phys. Rev. D \textbf{101}, no.11, 115015 (2020)}
[arXiv:2004.01096 [hep-ph]].

\bibitem{Altmannshofer:2020ywf}
W.~Altmannshofer, S.~Gori, H.~H.~Patel, S.~Profumo and D.~Tuckler,
\href{http://dx.doi.org/doi:10.1007/JHEP05(2020)069}{
JHEP \textbf{05}, 069 (2020)}
[arXiv:2002.01400 [hep-ph]].

\bibitem{Dev:2020qet}
P.~S.~Bhupal Dev, R.~Mohanta, S.~Patra and S.~Sahoo,
\href{http://dx.doi.org/doi:10.1103/PhysRevD.102.095012}{
Phys. Rev. D \textbf{102}, no.9, 095012 (2020)}
[arXiv:2004.09464 [hep-ph]].

\bibitem{Mecaj:2020opd}
B.~Mecaj and M.~Neubert,
[arXiv:2012.02186 [hep-ph]].

\bibitem{Hati:2020cyn}
C.~Hati, J.~Kriewald, J.~Orloff and A.~M.~Teixeira,
[arXiv:2012.05883 [hep-ph]].

\bibitem{Bhaskar:2021pml}
A.~Bhaskar, D.~Das, T.~Mandal, S.~Mitra and C.~Neeraj,
[arXiv:2101.12069 [hep-ph]].

\bibitem{Crivellin:2021egp}
A.~Crivellin, D.~M\"uller and L.~Schnell,
[arXiv:2101.07811 [hep-ph]].

\bibitem{Assad:2017iib}
N.~Assad, B.~Fornal and B.~Grinstein,
\href{http://dx.doi.org/doi:10.1016/j.physletb.2017.12.042}{
Phys. Lett. B \textbf{777} (2018), 324-331}
[arXiv:1708.06350 [hep-ph]].

\bibitem{Fornal:2018dqn}
B.~Fornal, S.~A.~Gadam and B.~Grinstein,
\href{http://dx.doi.org/doi:10.1103/PhysRevD.99.055025}{
Phys. Rev. D \textbf{99}, no.5, 055025 (2019)}
[arXiv:1812.01603 [hep-ph]].

\bibitem{DiLuzio:2019jyq}
L.~Di Luzio, M.~Kirk, A.~Lenz and T.~Rauh,
\href{http://dx.doi.org/doi:10.1007/JHEP12(2019)009}{
JHEP \textbf{12}, 009 (2019)}
[arXiv:1909.11087 [hep-ph]].

\bibitem{Altmannshofer:2021qrr}
W.~Altmannshofer and P.~Stangl,
[arXiv:2103.13370 [hep-ph]].

\bibitem{Glattauer:2015teq}
R.~Glattauer \textit{et al.} [Belle],
\href{http://dx.doi.org/doi:10.1103/PhysRevD.93.032006}{
Phys. Rev. D \textbf{93} (2016) no.3, 032006}
[arXiv:1510.03657 [hep-ex]].

\bibitem{DiLuzio:2017chi}
L.~Di Luzio and M.~Nardecchia,
\href{http://dx.doi.org/doi:10.1140/epjc/s10052-017-5118-9}{
Eur. Phys. J. C \textbf{77}, no.8, 536 (2017)}
[arXiv:1706.01868 [hep-ph]].


\bibitem{Fajfer:2012jt}
S.~Fajfer, J.~F.~Kamenik, I.~Nisandzic and J.~Zupan,
\href{http://dx.doi.org/doi:10.1103/PhysRevLett.109.161801}{
Phys. Rev. Lett. \textbf{109}, 161801 (2012)}
[arXiv:1206.1872 [hep-ph]].

\bibitem{Diaz:2017lit}
B.~Diaz, M.~Schmaltz and Y.~M.~Zhong,
\href{http://dx.doi.org/doi:10.1007/JHEP10(2017)097}{
JHEP \textbf{10}, 097 (2017)}
[arXiv:1706.05033 [hep-ph]].

\bibitem{Faroughy:2016osc}
D.~A.~Faroughy, A.~Greljo and J.~F.~Kamenik,
\href{http://dx.doi.org/doi:10.1016/j.physletb.2016.11.011}{
Phys. Lett. B \textbf{764}, 126-134 (2017)}
[arXiv:1609.07138 [hep-ph]].

\bibitem{Greljo:2017vvb}
A.~Greljo and D.~Marzocca,
\href{http://dx.doi.org/doi:10.1140/epjc/s10052-017-5119-8}{
Eur. Phys. J. C \textbf{77}, no.8, 548 (2017)}
[arXiv:1704.09015 [hep-ph]].

\bibitem{Aaboud:2017buh}
M.~Aaboud \textit{et al.} [ATLAS],
\href{http://dx.doi.org/doi:10.1007/JHEP10(2017)182}{
JHEP \textbf{10}, 182 (2017)}
[arXiv:1707.02424 [hep-ex]].

\bibitem{Sirunyan:2018ipj}
A.~M.~Sirunyan \textit{et al.} [CMS],
\href{http://dx.doi.org/doi:10.1007/JHEP04(2019)114}{
JHEP \textbf{04}, 114 (2019)}
[arXiv:1812.10443 [hep-ex]].

\bibitem{Alwall:2014hca}
J.~Alwall, R.~Frederix, S.~Frixione, V.~Hirschi, F.~Maltoni, O.~Mattelaer, H.~S.~Shao, T.~Stelzer, P.~Torrielli and M.~Zaro,
\href{http://dx.doi.org/doi:10.1007/JHEP07(2014)079}{
JHEP \textbf{07}, 079 (2014)}
[arXiv:1405.0301 [hep-ph]].

\bibitem{Degrande:2011ua}
C.~Degrande, C.~Duhr, B.~Fuks, D.~Grellscheid, O.~Mattelaer and T.~Reiter,
\href{http://dx.doi.org/doi:10.1016/j.cpc.2012.01.022}{
Comput. Phys. Commun. \textbf{183}, 1201-1214 (2012)}
[arXiv:1108.2040 [hep-ph]].

\bibitem{Alloul:2013bka}
A.~Alloul, N.~D.~Christensen, C.~Degrande, C.~Duhr and B.~Fuks,
\href{http://dx.doi.org/doi:10.1016/j.cpc.2014.04.012}{
Comput. Phys. Commun. \textbf{185}, 2250-2300 (2014)}
[arXiv:1310.1921 [hep-ph]].

\bibitem{Buckley:2014ana}
A.~Buckley, J.~Ferrando, S.~Lloyd, K.~Nordstr\"om, B.~Page, M.~R\"ufenacht, M.~Sch\"onherr and G.~Watt,
\href{http://dx.doi.org/doi:10.1140/epjc/s10052-015-3318-8}{
Eur. Phys. J. C \textbf{75}, 132 (2015)}
[arXiv:1412.7420 [hep-ph]].

\bibitem{Dorsner:2018ynv}
I.~Dor\v{s}ner and A.~Greljo,
\href{http://dx.doi.org/doi:10.1007/JHEP05(2018)126}{
JHEP \textbf{05}, 126 (2018)}
[arXiv:1801.07641 [hep-ph]].

\bibitem{Shtabovenko:2020gxv}
V.~Shtabovenko, R.~Mertig and F.~Orellana,
\href{http://dx.doi.org/doi:10.1016/j.cpc.2020.107478}{
Comput. Phys. Commun. \textbf{256}, 107478 (2020)}
[arXiv:2001.04407 [hep-ph]].

\bibitem{Shtabovenko:2016sxi}
V.~Shtabovenko, R.~Mertig and F.~Orellana,
\href{http://dx.doi.org/doi:10.1016/j.cpc.2016.06.008}{
Comput. Phys. Commun. \textbf{207}, 432-444 (2016)}
[arXiv:1601.01167 [hep-ph]].

\bibitem{Manohar:2016nzj}
A.~Manohar, P.~Nason, G.~P.~Salam and G.~Zanderighi,
\href{http://dx.doi.org/doi:10.1103/PhysRevLett.117.242002}{
Phys. Rev. Lett. \textbf{117}, no.24, 242002 (2016)}
[arXiv:1607.04266 [hep-ph]].

\bibitem{Manohar:2017eqh}
A.~V.~Manohar, P.~Nason, G.~P.~Salam and G.~Zanderighi,
\href{http://dx.doi.org/doi:10.1007/JHEP12(2017)046}{
JHEP \textbf{12}, 046 (2017)}
[arXiv:1708.01256 [hep-ph]].

\bibitem{Buonocore:2020nai}
L.~Buonocore, P.~Nason, F.~Tramontano and G.~Zanderighi,
\href{http://dx.doi.org/doi:10.1007/JHEP08(2020)019}{
JHEP \textbf{08}, no.08, 019 (2020)}
[arXiv:2005.06477 [hep-ph]].

\bibitem{Buonocore:2020erb}
L.~Buonocore, U.~Haisch, P.~Nason, F.~Tramontano and G.~Zanderighi,
\href{http://dx.doi.org/doi:10.1103/PhysRevLett.125.231804}{
Phys. Rev. Lett. \textbf{125}, no.23, 231804 (2020)}
[arXiv:2005.06475 [hep-ph]].

\bibitem{Greljo:2020tgv}
A.~Greljo and N.~Selimovic,
[arXiv:2012.02092 [hep-ph]].

\bibitem{Clark:2016jgm}
D.~B.~Clark, E.~Godat and F.~I.~Olness,
\href{http://dx.doi.org/doi:10.1016/j.cpc.2017.03.004}{
Comput. Phys. Commun. \textbf{216}, 126-137 (2017)}
[arXiv:1605.08012 [hep-ph]].

\bibitem{Haisch:2020xjd}
U.~Haisch and G.~Polesello,
[arXiv:2012.11474 [hep-ph]].

\bibitem{Thamm:2015zwa}
A.~Thamm, R.~Torre and A.~Wulzer,
\href{http://dx.doi.org/doi:10.1007/JHEP07(2015)100}{
JHEP \textbf{07}, 100 (2015)}
[arXiv:1502.01701 [hep-ph]].

\bibitem{Allanach:2017bta}
B.~C.~Allanach, B.~Gripaios and T.~You,
\href{http://dx.doi.org/doi:10.1007/JHEP03(2018)021}{
JHEP \textbf{03}, 021 (2018)}
[arXiv:1710.06363 [hep-ph]].

\bibitem{Martin:2009iq}
A.~D.~Martin, W.~J.~Stirling, R.~S.~Thorne and G.~Watt,
\href{http://dx.doi.org/doi:10.1140/epjc/s10052-009-1072-5}{
Eur. Phys. J. C \textbf{63}, 189-285 (2009)}
[arXiv:0901.0002 [hep-ph]].

\bibitem{Ball:2013hta}
R.~D.~Ball \textit{et al.} [NNPDF],
\href{http://dx.doi.org/doi:10.1016/j.nuclphysb.2013.10.010}{
Nucl. Phys. B \textbf{877}, 290-320 (2013)}
[arXiv:1308.0598 [hep-ph]].

\bibitem{Allanach:2019zfr}
B.~C.~Allanach, T.~Corbett and M.~Madigan,
\href{http://dx.doi.org/doi:10.1140/epjc/s10052-020-7722-3}{
Eur. Phys. J. C \textbf{80}, no.2, 170 (2020)}
[arXiv:1911.04455 [hep-ph]].


\end{thebibliography}
\end{document}